\documentclass[12pt,preprint]{aastex}
\newcommand{\beq}{\begin{equation}}
\newcommand{\eeq}{\end{equation}}

\newcommand{\ag}{\mbox{ \raisebox{-.4ex}{$\stackrel{\textstyle >}{\sim}$} }}
\newcommand{\al}{\mbox{ \raisebox{-.4ex}{$\stackrel{\textstyle <}{\sim}$} }}

\newcommand{\cicol}{N_{\rm C\,I}}
\newcommand{\ciicol}{N_{\rm C\,II}}
\newcommand{\hicol}{N_{\rm H\,I}}

\slugcomment{Submitted for Publication in the Astrophysical Journal}

\shortauthors{Wolfire et al.}
\shorttitle{}
\begin{document}

\title{Chemical Rates on Small Grains and PAHs: ${\rm C^+}$ Recombination 
and  ${\rm H_2}$ Formation}

\author{
Mark\ G. Wolfire}
\affil{Department of Astronomy, University of Maryland,
College Park, MD 20742-2421}
\email{mwolfire@astro.umd.edu}

\author{
A. G. G. M. Tielens}
\affil{Space Science and Astrobiology Division, NASA Ames Research
  Center, 
MS 245-3, Moffett Field,
 CA 94035}
\email{atielens@mail.arc.nasa.gov}

\author{
David Hollenbach}
\affil{
Space Science and Astrobiology Division, NASA Ames Research
  Center, 
MS 245-3, Moffett Field,
 CA 94035}
\email{hollenbach@ism.arc.nasa.gov}

\and
\author{
M. J. Kaufman}
\affil{Department of Physics, San Jose State University,
One Washington Square, San Jose, CA 95192-0106 
}
\email{mkaufman@email.sjsu.edu}


\begin{abstract}

We use observations of the \ion{C}{1}, \ion{C}{2}, \ion{H}{1}, and 
${\rm H_2}$ column densities
along lines of sight in the Galactic plane to determine the formation rate 
of ${\rm H_2}$ on grains and to determine chemical reaction rates 
with Polycyclic 
Aromatic Hydrocarbons. Photodissociation region models
are used to find the best 
fit parameters to the observed columns. 
We find the  ${\rm H_2}$ formation rate on grains
has a low rate ($R \sim 1\times 10^{-17}$ ${\rm cm^{3}}$ ${\rm s^{-1}}$) 
along lines 
of sight with low column density ($A_{\rm V} \al 0.25$) and low molecular 
fraction
($f_{\rm H_2} \al 10^{-4}$). At higher column densities 
($0.25 \le A_{\rm V} \le 2.13$),
we find a rate
of $R\sim 3.5 \times 10^{-17}$ ${\rm cm^{3}}$ ${\rm s^{-1}}$. The 
lower rate at low
column densities could be the result of grain processing by 
interstellar shocks which may deplete the grain surface area or
process the sites of ${\rm H +H}$ formation, thereby inhibiting 
${\rm H_2}$ production.
Alternatively, the formation rate  may be normal, and the 
low molecular fraction may be the result of lines of 
sight which
graze larger clouds. Such lines of sight would have a reduced
${\rm H_2}$ self-shielding compared to the line-of-sight column. 
We find the reaction 
${\rm C^+} + {\rm PAH^-} \rightarrow {\rm C} + {\rm PAH^0}$ is best fit
with a rate $2.4\times 10^{-7}\Phi_{\rm PAH} T_2^{-0.5}$ ${\rm cm^{3}}$
${\rm s^{-1}}$ with $T_2= T/100$ K and
the reaction 
${\rm C^+} + {\rm PAH^0} \rightarrow {\rm C} + {\rm PAH^+}$ is best fit
with a rate $8.8\times 10^{-9}\Phi_{\rm PAH}$ ${\rm cm^{3}}$
${\rm s^{-1}}$. In high column density gas we find $\Phi_{\rm PAH} \sim 0.4$.
In low column density gas, $\Phi_{\rm PAH}$ is less well constrained with
$\Phi_{\rm PAH} \sim 0.2 - 0.4$.

\end{abstract}


\section{INTRODUCTION}
\label{sec:introduction}

Much of the chemistry in the interstellar medium (ISM) proceeds via
two body gas phase reactions; however,
based on reaction time arguments it is well established that
the formation of ${\rm H_2}$  must proceed predominantly
by grain surface reactions \citep{hollenbach1971} and thus
the equilibrium abundance of  ${\rm H_2}$ in diffuse gas depends on
the balance between dissociation from ultraviolet radiation 
and the formation on  grains. 
In addition, \cite{lepp1988}  suggested that 
in diffuse clouds,
reactions with small grains or large molecules can affect the charge
balance and abundance of metal ions. More recent theoretical
\citep[e.g.,][]{bakes1998,weingartner2001,wolfire2003,lips2003, chang2006} and
laboratory, \citep{cazaux2004} investigations have
explored the rates of ${\rm H_2}$ formation and 
chemical reactions on surfaces of small grains
and their dependence on environmental factors. 
The chemical rates have been difficult to pin down
in part because the exact abundance, type, surface area, and physical
properties 
of the grains/large molecules are not well determined.

Studies with a more observational approach
have also estimated ${\rm H_2}$ formation
rates and chemical rates with grains
\citep[e.g.,][]{jura1975,browning2003,welty2003}.
\cite{jura1975} estimated a formation
rate of about $R~\sim 3\times 10^{-17}$ ${\rm cm^{3}}$ ${\rm s^{-1}}$ in
diffuse gas of low column density; however, 
one could see in the {\em Copernicus} survey of \cite{savage1977} 
that a single ${\rm H_2}$ formation
rate can not explain all of the observed column densities
even accounting for the 
expected variation in ultraviolet radiation field and gas density. 
A more recent analysis by \cite{gry2002} using data from the {\em Far
Ultraviolet Spectroscopic Explorer} (FUSE) find a formation
rate of $R\sim 4\times 10^{-17}$ ${\rm cm^{3}}$ ${\rm s^{-1}}$ in
diffuse gas and \cite{browning2003} estimates the rate at 2.5 - 10 
times lower in the low metallicity environments of the Large and Small
Magellanic Clouds. 
\cite{habart2004} suggests that the ${\rm H_2}$
formation rate varies in molecular clouds exposed to intense
radiation by a factor of about five, perhaps due to the temperature 
dependence of the formation process on grain surfaces \citep{cazaux2004}.
In this paper we take 
a semi-empirical approach 
by allowing the ensemble of recent observations to guide us in the proper 
``astrophysical'' rates rather than use rates based on 
laboratory measurements with an assumed grain type and distribution. 

\cite{bakes1998} investigated the effects of Polycyclic Aromatic
Hydrocarbons (PAHs) on 
the chemistry of dense regions exposed to intense ultraviolet
radiation. They found that charge exchange reactions with PAHs are
important in setting the column density of neutral metals, in
particular \ion{C}{1}.  \cite{welty2003} found
patterns of neutral column densities in diffuse clouds higher 
than suggested by 
electron recombination alone and suggested that 
recombination on grains might
be important. \cite{weingartner2001} carried out a detailed analysis
of the ``grain assisted'' rates required to explain the observed
column densities  along the line of sight to 
23 Ori. \cite{wolfire1995,wolfire2003} found that reactions with PAHs
can affect the ionization balance of metals and also atomic hydrogen,
in diffuse gas, in particular in the warm neutral medium. This
has an important effect on the grain charge and thus the heating
produced by the grain photoelectric effect; the dominant heating
process in the diffuse ISM.  

In \cite{wolfire2003} we adopted rates with PAHs based 
on the \cite{draine1987}
formalism for interactions with small grains, but modified 
the rates to match a preliminary assessment of the 
observed $\cicol/\ciicol$ ratio in diffuse gas. This ratio is a 
particularly good
diagnostic because carbon is abundant, the rates of
${\rm C^+}$ recombination on PAH anions effectively compete with
electron recombination, and carbon exists
in the diffuse ISM outside of \ion{H}{2} regions
as either \ion{C}{1} or \ion{C}{2}
but not higher ionization stages.

For several years, the {\em Copernicus} column densities of \ion{H}{1}
\citep{bohlin1978},
${\rm H_2}$ \citep{savage1977}, \ion{C}{1} \citep{jenkins1979,jenkins1983},
and \ion{C}{2} \citep{hobbs1982}
were the only data available.
More recently the combined observations from 
the {\em Hubble Space
  Telescope} Goddard High Resolution Spectrograph (GHRS) and 
Space Telescope Imaging Spectrograph (STIS) instruments along with
the {\em Far Ultraviolet Spectroscopic Explorer} (FUSE) have produced
many more sight lines to study in particular in ${\rm H_2}$ 
\citep[e.g.,][]{rachford2002, cartledge2004}, \ion{C}{1}
\citep[e.g.,][]{zsargo1997,zsargo2003}, and \ion{C}{2} 
\citep[e.g.,][]{sofia2004}.
In \S~2 we present the observations
used in this study and in \S~3 we discuss our cloud models and our 
fitting procedure. In \S~4 we present our results
for the  best fit rates for grain reactions and for ${\rm H_2}$ formation.
We also discuss some of the interesting sight lines which seem to
have outlying points. We summarize our main conclusions in \S~5.

\section{OBSERVATIONS}
\label{sec:observations}

Observations of the \ion{H}{1}, ${\rm H_2}$, \ion{C}{1}, and 
\ion{C}{2} columns are listed in Table~\ref{tbl:observations}
and are taken from the literature as noted. 
We have included only those lines of sight which have measured values of 
$\log \hicol$, $\log N_{\rm H_2}$, and $\log \cicol$, and
have not included lines of sight with upper or lower limits in these
quantities.  The \ion{C}{2} column
is measured directly only towards thirteen sources  \citep{sofia2004}. 
Six of these have definite \ion{C}{1} columns and are included in our sample.
For all other sources we use the conversion 
  $\ciicol ={\cal A}_{\rm C}\times N$ where 
$N\equiv \hicol + 2N_{\rm H_2}$, and
${\cal A}_{\rm C}=1.6\times 10^{-4}$ is the mean gas phase abundance of carbon 
per hydrogen nucleus
found by \cite{sofia2004} from all measured values of the \ion{C}{2}
carbon abundance in the diffuse ISM. 

Most of the \ion{H}{1} and ${\rm H_2}$ columns and uncertainties 
are from the {\em Copernicus} ultraviolet absorption 
spectroscopy observations of \cite{bohlin1978}
and \cite{savage1977}. More recently, 
\cite{cartledge2004} updated
several of these lines of sight with FUSE observations. 
The columns are in good
agreement with the previous results, but with smaller quoted 
uncertainties. In cases of overlap we use the \cite{cartledge2004}
results. Additional ${\rm H_2}$ columns are provided by
FUSE surveys \citep[e.g.,][]{rachford2002}.
In several low column density lines of sight \cite{savage1977} did
not quote an uncertainty for the ${\rm H_2}$ columns. 
In these cases we use the uncertainty $\sigma(\log N_{\rm H_2}) = 0.20$
as suggested
by D.\ Welty\footnote{http://astro.uchicago.edu/home/web/welty/coldens.html}
and is comparable to the maximum quoted value for all observations.

Many of the $\cicol$ columns are from the {\em Copernicus} ultraviolet
absorption observations
of \cite{jenkins1979} and \cite{jenkins1983}. Additional \ion{C}{1} columns
are provided for example, by GHRS observations and FUSE 
\citep{zsargo1997,zsargo2003,jt2001}.
The method used by \cite{jt2001} to determine \ion{C}{1} columns
  does not yield a direct measure of the random errors in the observed
  velocity integrated column density. We adopt an uncertainty comparable to 
  similar STIS and {\em FUSE}
  observations at similar column densities \citep[e.g.,][]{sonn2002}.

\section{MODELS}
\label{sec:models}  

\subsection{Reaction Rates with PAHs and Carbon Chemistry}
\label{subsec:PAH rates}

The PAH rates in \cite{wolfire2003} were calculated using the  equations
given by \cite{draine1987} for a disk shaped PAH with the number of Carbon
atoms $N_{\rm C} = 35$ and a PAH abundance  of $6\times 10^{-7}$ per
hydrogen nucleus. \cite{wolfire2003} found it necessary to modify the rates 
by a factor $\Phi_{\rm PAH}$ to match a preliminary assessment of the 
observed $\cicol/\ciicol$ ratio in the diffuse ISM. The factor 
$\Phi_{\rm PAH}$ is
to include a wide range of unknowns in the rates including 
the PAH size, geometry, and abundance. Although the full \cite{draine1987}
formalism is used in our code, over typical temperatures and densities
found in the diffuse ISM the rates can be closely approximated by
simple formulae. For ${\rm C^+}$ recombination on PAHs we have
\beq
{\rm C^+ + PAH^-} \rightarrow {\rm C + PAH^{0}}, \,\,\,\, 
            k_1 = 2.4\times 10^{-7} \Phi_{\rm PAH} T_2^{-0.5}\,\,\, 
           {\rm cm^3\,\, s^{-1}}\, , 
\label{eq:cprecompahm}
\eeq
where $T_2 \equiv T/(100\,\,{\rm K})$,
and for charge exchange
\beq
{\rm C^+ + PAH^0} \rightarrow {\rm C + PAH^{+}}, \,\,\,\, 
            k_2 = 8.8\times 10^{-9} \Phi_{\rm PAH}\,\,\, 
           {\rm cm^3\,\, s^{-1}}\, . 
\label{eq:cprecompahn}
\eeq
Several rates were given 
in \cite{wolfire2003} Appendix C2, mainly related to the charge balance for
hydrogen and the electron abundance.  We provide a summary of our 
rate set in Appendix \ref{appen:pahrates}.
  
In addition to reactions with PAHs, ${\rm C^+}$ can also recombine with 
electrons in the gas phase via radiative or dielectronic recombination.
It has recently come to our attention that dielectronic recombination
of ${\rm C^+}$ can be significant in diffuse gas 
(G. Ferland, private communication). 
The total (dielectronic plus radiative) rates by \cite{altun2004} and 
those posted on-line for lower temperatures
by Badnell ( http://amdpp.phys.strath.ac.uk/tamoc/DATA/) are higher
by a factor of $\sim 2$ at 100 K than rates by \cite{nahar1997}. 
For the majority of this paper we use the Badnell rates, but discuss
the implications if they are lower as suggested by the previous data.
A fit to the  total rate is given by 
\beq
     {\rm C^+}\,\, + e \rightarrow {\rm C^0} \,\,\,\, k_3 = 1.8\times 10^{-11}
     T_2^{-0.83} \,\,\, {\rm cm^3\,\, s^{-1}}\, ,
\label{eq:cprecome}
\eeq
where the fit is good to within 8\% for 20 K $ < T < 300$ K.

The ${\rm C^+}$ is formed mainly by the photoionization of C by the 
interstellar radiation field
\beq     
{\rm C\,\, + }\, h\nu \rightarrow {\rm C^+\,\, +}\,\,  e, \,\,\,\,
    k_4 = 2.1\times 10^{-10} G_0 \exp (-2.6A_{\rm V})\,\,\, {\rm s^{-1}}
\label{eq:c1ionization}
\eeq 
where $G_0$
 is the FUV (6 eV $\le h\nu \le 13.6$ eV) interstellar radiation 
field measured in units of the  \cite{habing1968} field
($G_0 = 1$ is $1.3\times 10^{-4}$ erg ${\rm cm^{-2}}$ ${\rm s^{-1}}$ 
${\rm sr^{-1}}$).
In \cite{wolfire2003}, we adopted $G_0=1.7$, a field strength 
comparable to the 
\cite{draine1978} interstellar field. Balancing formation and
destruction processes, in clouds optically thin to the FUV radiation,
the $\cicol/\ciicol$ ratio is mainly a function of $n\Phi_{\rm PAH}/G_0$.

\subsection{${\rm H_2}$ Formation}
\label{subsec:H2formation}

Molecular hydrogen formation in diffuse gas proceeds via reactions with 
atomic hydrogen on grain surfaces. We adopt a formation rate per unit
volume which
goes as $n n_{\rm H\,I} R$ where $n$ is the hydrogen nucleus density,
$n_{\rm H\, I}$ is the density of atomic hydrogen, and $R$ is
a constant of order $3\times 10^{-17}$ ${\rm cm^{3}}$ 
${\rm s^{-1}}$. The constant includes the surface area
of grains per hydrogen nucleus. (If the metallicity were to vary 
as $Z$, and the grain surface area per H proportional to $Z$, then the rate
$R$ would go as $Z$.) Previous investigators have included both
a gas and grain temperature
dependence in this rate equation intended to include the collision
rate of atomic hydrogen with grains, 
the sticking coefficient
of hydrogen on grains (which is a function of both the hydrogen
temperature and grain temperature), and the thermal diffusion 
of atoms across grain surfaces and thermal evaporation from these
surfaces (which is a function of the grain temperature).
The collision rate increases with
temperature while the sticking coefficient and formation efficiency
decrease with temperature \citep{hollenbach1971,burke1983,pirronello1999,
cazaux2004}. In light 
of the uncertainties of the details of the various processes, 
 \cite{kaufman1999} adopted an average rate which was independent of 
temperature.

The dissociation rate of ${\rm H_2}$ per unit volume goes as   
\beq
G_0 I n_{\rm H_2} \beta (N_{\rm H_2})\exp(-2.5A_{\rm V})  
\label{eq:h2dissociation}
\eeq
where $I$ is the unshielded
photodissociation rate in the local interstellar field ($G_0=1$), 
and $\beta (N_{\rm H_2})$ is the ${\rm H_2}$ self-shielding
factor. We take $I = 4.7\times 10^{-11}$ ${\rm s^{-1}}$ 
from \cite{abgrall1992}.
The unshielded rate could change depending on the 
detailed population of the molecular hydrogen rotational and
vibrational levels resulting from different
incident FUV fields, 
gas densities, and temperatures. 
We have compared our constant unshielded dissociation 
rate $I$, with results produced with 
the code available on-line from the 
Meudon Group\footnote{See http://aristote.obspm.fr/MIS}.
As discussed in \cite{lepetit2006} they include a more detailed 
treatment of ${\rm H_2}$ dissociation and shielding. 
We ran their
code at $n=30$ and 100 ${\rm cm^{-3}}$, $G_0 = 1.7$, 
and $A_{\rm V} = 0.1$, 0.3, and 1, and found good agreement
with less than a 10\% difference in values of $I$. 
We use the self-shielding factor from 
\cite{draine1996} (their equation 37) which is a fit to their detailed
${\rm H_2}$ destruction calculation. The equilibrium 
molecular fraction $2n_{\rm H_2}/(n_{\rm H\,I} + 2n_{\rm H_2})$, 
for unshielded ${\rm H_2}$ ($\beta = 1$) equals 
$2nR/G_0I$. For constant $I$, the molecular fraction is proportional
to  $2nR/G_0$. Thus, to first order and ignoring shielding by dust
or ${\rm H_2}$, the 
$\cicol/\ciicol$ ratio depends on $\Phi_{\rm PAH}n/G_0$ while the
$N_{\rm H_2}/\hicol$ ratio depends on $Rn/G_0$.

\subsection{2-Sided Models}
\label{subsec:2sidedmodels}

We see from sections  \ref{subsec:PAH rates} and
\ref{subsec:H2formation} that the $\cicol/\ciicol$ ratio 
and the ${\rm H_2}$ column $N_{\rm H_2}$ depends on 
$\Phi_{\rm PAH}$, $R$, and the ratio of $G_0/n$. 
The total column density $N= \hicol + 2N_{\rm H_2}$ also enters
due to the extinction of the incident FUV field 
in equations (\ref{eq:c1ionization}) and (\ref{eq:h2dissociation})
 and also in the self-shielding
factor $\beta$ (which depends on the ${\rm H_2}$ column density).
The total column density is known for each source. Thus we
can construct models for each source by holding the total column
density fixed, and varying $G_0/n$, $\Phi_{\rm PAH}$, and $R$, until
good fits are obtained to the observed columns densities, 
$N_{\rm C\,I}/N_{\rm C\,II}$ and $2N_{\rm H_2}/N$ 

Since many of the observations have cloud columns  $A_{\rm V}\al 1$ the effects
of radiation incident on the near and far side of the cloud need to 
be included. This radiation affects both the photoionization of C and
the dissociation of ${\rm H_2}$. We have modified the 1-sided 
Photodissociation Region (PDR) code
of \cite{kaufman1999} to include the 2-sided incident radiation. 
The code calculates the equilibrium chemical abundances and gas
temperature in a gas layer exposed to X-rays and FUV radiation. 
In the one-sided case, calculation of the chemistry, cooling, and line 
transfer proceeds from the surface to the cloud center in a single pass 
since these parameters depend only on the cloud properties closer to the 
surface. 
In the two-sided case, at a given point in the cloud,
the optical depth in the cooling lines and the ${\rm H_2}$ column
is required towards both surfaces, thus an iterative procedure is required.
We first calculate the structure of a cloud illuminated from one side 
to a depth of one-half the visual extinction of the desired two-sided model. 
This provides us with initial estimates of line optical depths as well 
as shielding columns of ${\rm H_2}$ and CO. We then iterate the
chemical abundance profiles, optical depths, and shielding factors.

We have further modified the physics and chemistry of the 
\cite{kaufman1999} code according to the discussion in 
\cite{wolfire2003} who assumed a higher abundance of PAHs and
a slower rate of interaction of ions and PAHs. 
These changes largely offset each other, with a resulting minor effect 
on the grain heating rates and ion chemistry. In addition, we have adopted 
the results of \cite{Pequignot1990} for the rate coefficient for collisional 
excitation of O I by H, and those of \cite{mccall2003} for the ${\rm H_3^+}$ 
dissociative recombination rate coefficient.
We have also included the illumination of the cloud by the interstellar 
soft X-ray 
radiation field as discussed in \cite{wolfire2003}.
Results from this 2-sided code have been previously presented in
\cite{neufeld2005}, and \cite{snow2006}.

We have compared our calculated ${\rm H_2}$ column densities with
those produced by the Meudon code. For the same model parameters in 
section \ref{subsec:H2formation}, and running the code in 2-sided mode,
we find that our ${\rm H_2}$ columns 
agree to within a factor of 2 (with our columns lower) 
and with the largest difference at low column densities 
($A_{\rm V}\sim 0.1$). The agreement is within the typical 
observational error. 

\subsection{Fitting $\Phi_{\rm PAH}$ and $R$}

We determine the best fit $\Phi_{\rm PAH}$ and $R$ by computing
a grid of models consisting of 12 values of $\Phi_{\rm PAH}$ 
(0.01, 0.1, 0.2 ,0.3,
0.4, 0.5, 0.6, 0.7, 0.8, 0.9, 1.0, 1.1)
and 11 values of $R$ (0.5, 1, 2, 3, 4, 5, 6, 7, 8, 9,
10) $\times 10^{-17}$ ${\rm cm^{3}}$ ${\rm s^{-1}}$. At each grid point,
(i.e., for fixed $\Phi_{\rm PAH}$ and $R$), we 
find the best fit $G_0/n$ values for each source by minimizing the 
$\chi^2_s$
\beq
   \chi^2_s(\Phi_{\rm PAH}, R) = 
\left \{ \frac{\log f_{\rm H_2}^{\rm model} - \log f_{\rm H_2}}
       { \sigma[\log f_{\rm H_2} ]} \right \}^2 +
      \left \{  \frac{\log f_{\rm C\,I}^{\rm model} - \log f_{\rm C\,I}}
          {\sigma [ \log f_{\rm C\,I} ]} \right \}^2 ,
\eeq
where $\chi^2_s$ is the $\chi^2$ value for source $s$, 
$f_{\rm H_2}$ and $f_{\rm C\,I}$ are the
observed values for the column density ratios
$f_{\rm H_2} = 2N_{\rm H_2}/[\hicol + 2N_{\rm H_2}]$,
$f_{\rm C\,I} = {\cicol/\ciicol}$,
$f_{\rm H_2}^{\rm model}$, and $f_{\rm C\,I}^{\rm model}$ are the
calculated values of $f_{\rm H_2}$ and $f_{\rm C\,I}$ (dependent
upon $N$ -- which is measured, and on the unknowns
$\Phi_{\rm PAH}$, $R$, and $G_0/n$).

We will calculate the  $\chi^2$ value over a subset of the sources
within a range of $A_{\rm V}$ values.  Thus
the reduced $\chi^2$ value at grid point ($\Phi_{\rm PAH}, R)$ is the sum
over sources within the restricted $A_{\rm V}$ bin:
\beq
   \chi^2(\Phi_{\rm PAH}, R) = \sum_s^{\rm N_s} \chi^2_s/(N_s-2) .
\eeq
where $N_s$ is the total number of sources in the bin.

We find the uncertainty in the measurements 
$\sigma (\log f_{\rm H_2})$ and $\sigma (\log f_{\rm C\,I})$
using the quoted uncertainties
$\sigma(\log \hicol)$,  $\sigma(\log N_{\rm H_2})$, $\sigma(\log \cicol)$,
for the \ion{H}{1}, ${\rm H_2}$, and \ion{C}{1} columns, along with
the standard propagation of error formula \citep[e.g.,][]{taylor1997}.  
Thus
\beq
\sigma({\log f_{\rm H_2}}) = \left [ \sigma( \log N_{\rm H_2})^2 +
          \sigma( \log N)^2 \right ]^{1/2},
\label{eq:sigmafh2}
\eeq
where $\sigma( \log N)$ is the uncertainty in the total column density,
\beq
\sigma(\log N) = \left [ \sigma( \log N_{\rm H_2})^2 f_{\rm H_2}^2+
          \sigma( \log \hicol)^2 f_{{\rm H\,I}}^2 \right ]^{1/2},
\label{eq:sigman}
\eeq
and $f_{\rm H\,I}$ is the fraction of atomic hydrogen
$f_{\rm H\,I}=\hicol /[\hicol + 2N_{\rm H_2}]$. For observed \ion{C}{2}
column densities,
the uncertainty in $\log f_{\rm C\,I}$ is given by
\beq
  \sigma( \log f_{\rm C\,I} ) = \left [\sigma (\log \cicol)^2 +
                       \sigma (\log \ciicol)^2 \right ]^{1/2},
\label{eq:sigmacicii1}
\eeq 
while for estimated \ion{C}{2} column densities the uncertainty in
 $\log f_{\rm C\,I}$ is given by
\beq
  \sigma( \log f_{\rm C\,I} ) = \left [ \sigma (\log \cicol)^2 +
                       \sigma (\log N)^2 \right ]^{1/2}.
\label{eq:sigmacicii2}
\eeq 

In practice, for the $\chi^2$ test,  we set $G_0 = 1.7$ and vary $n$.
We estimate the initial range of $n$ values
from the analytic expressions in Appendix
\ref{appen:estimatinggn}. We run models
stepping in values of $n$ until we cross the minimum in
$\chi^2$. Then we apply a parabolic interpolation to find
the minimum $\chi^2$ and associated values of density and  
column densities.

\section{RESULTS and DISCUSSION}
\label{sec:resultsanddiscussion}

\subsection{Fixed Column Density Models}
\label{subsec:fixedrates}

We first carry out model runs for four fixed cloud column 
densities $A_{\rm V} = 0.1$, 0.5, 1.0, and 1.5 with 
variable $\Phi_{\rm PAH}$, $R$, and $G_0/n$. 
The data and model results are displayed in plots 
of $\cicol/\ciicol$ versus $f_{\rm H_2}$ with source column
densities grouped in four $A_{\rm V}$ 
bins. These plots are intended to show the range of
observed values, the effects of varying $\Phi_{\rm PAH}$ 
and $R$,
and an ``eye ball'' assessment of reasonable values. 
The $A_{\rm V}$ bins
span the range of observed $A_{\rm V}$ values with the 
smallest being $A_{\rm V} = 0.03$ and the  
largest $A_{\rm V} = 2.13$. The $A_{\rm V}$ bins have been chosen
to display natural groupings of the data which we further test
using the $\chi^2$ results.

We present in Figure \ref{fig:ratioav025} 
the observed and calculated 
ratio of $\cicol/\ciicol$ versus $f_{\rm H_2}$ for observations 
in the range
$0.03 \al A_{\rm V} \al 0.25$ and $N_{\rm H_2} \al 10^{17}$ 
cm$^{-2}$. 
Figure \ref{fig:ratioav025}
focuses on the low $f_{\rm H_2}$ ($\al 10^{-4}$) 
and low ${A_{\rm V}}$ directions.
The curves are for a model cloud column density of  
$A_{\rm V} = 0.1$, 
with $G_0 = 1.7$ and 5.1, and density points at
$n = 10$, 20, 30, and 40 ${\rm cm^{-3}}$ (with the
exception of two curves with three density points at
$n = 5$, 10, and 20 ${\rm cm^{-3}}$). 
Diamonds ($\Diamond$) indicate models with
$G_0 = 1.7$ and triangles ($\triangle$) indicate models
with $G_0 = 5.1$.
We have labeled the model curves with their corresponding
values of $\Phi_{\rm PAH}/R_{-17}$ where $R_{-17}=R/1\times 10^{-17}$.
The curves with 
$\Phi_{\rm PAH}/R_{-17} \le 1/3$ use $R = 3\times 10^{-17}$ ${\rm cm^{3}}$
${\rm s^{-1}}$ with values of $\Phi_{\rm PAH}=1$ 
($\Phi_{\rm PAH}/R_{-17} = 1/3$),
$\Phi_{\rm PAH}=0.5$ ($\Phi_{\rm PAH}/R_{-17} = 1/6$), and
$\Phi_{\rm PAH}=0.01$ ($\Phi_{\rm PAH}/R_{-17} = 1/300$). We have
added three additional curves with lower values of $R$ namely 
$R = 2\times 10^{-17}$ ${\rm cm^{3}}$ ${\rm s^{-1}}$, $\Phi_{\rm PAH} = 1$
($\Phi_{\rm PAH}/R_{-17} = 1/2$);
$R = 1\times 10^{-17}$ ${\rm cm^{3}}$ ${\rm s^{-1}}$, $\Phi_{\rm PAH} = 0.5$
($\Phi_{\rm PAH}/R_{-17} = 1/2$);  and 
$R = 1\times 10^{-17}$ ${\rm cm^{3}}$ ${\rm s^{-1}}$, $\Phi_{\rm PAH} = 1.0$
($\Phi_{\rm PAH}/R_{-17} = 1$) 
and one additional curve with higher values
of $\Phi_{\rm PAH}$ namely $R = 2\times 10^{-17}$ ${\rm cm^{3}}$ 
${\rm s^{-1}}$, $\Phi_{\rm PAH} = 2.0$ ($\Phi_{\rm PAH}/R_{-17} = 1$). 

  The curves
with the same value of $\Phi_{\rm PAH}/R_{-17}$, although not exactly the same,
differ by less than a factor of 1.5. 
We see that values of the ratio $\Phi_{\rm PAH}/R_{-17} \approx 1/2-1/3$ 
are required to  match the bulk of the observations. 
For $\Phi_{\rm PAH} \le 1$ this
requires $R\le 2 \times 10^{-17}$ ${\rm cm^{3}}$ ${\rm s^{-1}}$ and
an ${\rm H_2}$ formation rate smaller than the standard 
$3 \times 10^{-17}$ ${\rm cm^{3}}$ ${\rm s^{-1}}$ is necessary.


The subsequent Figures (\ref{fig:ratioav075}, \ref{fig:ratioav125},
and \ref{fig:ratioav213}) focus on the high molecular
abundance lines of sight ($f_{\rm H_2} > 10^{-3}$ and 
$N_{\rm H_2} > 10^{18}$ cm$^{-2}$). 
The observations are limited to the range 
$0.25 \al A_{\rm V} \al 0.75$ (Fig.\ [\ref{fig:ratioav075}]), 
$0.75 \al A_{\rm V} \al 1.25$ (Fig.\ [\ref{fig:ratioav125}]), and
$1.25 \al A_{\rm V} \al 2.13$ (Fig.\ [\ref{fig:ratioav213}])
and model curves are for $A_{\rm V} = 0.5$ (Fig.\ [\ref{fig:ratioav075}]), 
$A_{\rm V} = 1.0$ (Fig.\ [\ref{fig:ratioav125}]), and
$A_{\rm V} = 1.5$ (Fig.\ [\ref{fig:ratioav213}]).
The models all 
use $R=3\times 10^{-17}$ ${\rm cm^{3}}$ ${\rm s^{-1}}$
with values of ${\Phi_{\rm PAH}} = 0.01$, 0.5, and 1.0,
and density, $n = 10$, 30, 100, and
300 ${\rm cm^{-3}}$.
Comparing Figures \ref{fig:ratioav025} and \ref{fig:ratioav075} 
we see a large jump in the molecular fraction with
no observations at intermediate values. This jump is 
probably the result of the abrupt turn on of 
molecular hydrogen due to the effects of self-shielding.
Self-shielding of ${\rm H_2}$ at $N \ge 5\times 10^{20}$ 
${\rm cm^{-2}}$ ($A_{\rm V} \ge 0.25$) 
was seen in the \cite{savage1977} data
and also in the more recent FUSE surveys \citep{gillmon2006}.
In Figure \ref{fig:ratioav075},
two obvious outliers are 23 Ori ($A_{\rm V} = 0.28$)
at high $\cicol/\ciicol$ ratio ($= 1.1\times 10^{-2}$)
and $\pi$ Sco ($A_{\rm V} = 0.28$) at low
$\cicol/\ciicol$ ratio ($=1.2\times 10^{-3}$).
 These are  excluded from the $\chi^2$ tests and 
discussed further in  subsection~\ref{subsec:highav}.

 
The variation of the curves 
shown in Figures \ref{fig:ratioav025} 
through \ref{fig:ratioav213} as functions of  $G_0/n$, $\phi_{\rm PAH}$, 
and $R$
is discussed in Appendix \ref{appen:estimatinggn}.
Points with the same $G_0/(nR)$ and fixed column, or $A_{\rm V}$, 
have the same molecular fraction $f_{\rm H_2}$ independent of 
$\Phi_{\rm PAH}$. At the highest values of $G_0/n$ the molecular 
fraction is lowest due to rapid photodissociation of molecular hydrogen.
In the limit of no ${\rm H_2}$ self-shielding and no dust extinction,
the limiting molecular
fraction is given by $f_{\rm H_2} = {2nR}/({G_0 I})$. As $G_0/n$ decreases,
the molecular fraction increases with
a limiting value of $f_{\rm H_2} = 1$ for fully molecular gas. 
As the cloud column density (and thus $A_{\rm V}$) increases the local
FUV field drops due to dust extinction and the molecular fraction
rises due to a lower ${\rm H_2}$ photodissociation rate. In addition, 
the self-shielding by ${\rm H_2}$ raises the molecular fraction as the
column density increases.

The variation in the $\cicol/\ciicol$ ratio can be understood
in four limiting regimes: high/low $\Phi_{\rm PAH}$, and high/low $G_0/n$.
First consider the high $\Phi_{\rm PAH}$ ($0.5$ and $1.0$) curves.
For low $G_0/n$ the destruction of ${\rm C^+}$ is dominated
by recombination on ${\rm PAH^-}$ and 
$\cicol/\ciicol \propto \Phi_{\rm PAH}^2 (n/G_0)^2$ 
(See Appendix \ref{appen:variation}, eq.~[\ref{eq:Clowgn}]).
For high 
$G_0/n$, the destruction of ${\rm C^+}$ is dominated by charge
exchange with ${\rm PAH^0}$ and electron recombination and thus
$\cicol/\ciicol \propto \Phi_{\rm PAH} n/G_0$ (eq.~[\ref{eq:Chighgn}]).
In the case of low $\Phi_{\rm PAH}=0.01$ the
${\rm C^+}$ destruction is dominated by recombination with
free electrons $n_e$. For low $G_0/n$ the electrons are provided by 
photoionization of
C, and $\cicol/\ciicol \propto n/G_0$ (eq.~[\ref{eq:Clowgnlowphi}]).
For high $G_0/n$, i.e., low values of $n$, the electrons are provided
by the soft-X ray ionization of H, and 
$\cicol/\ciicol \propto n^{1/2}/G_0$ (eq.~[\ref{eq:Chighgnlowphi}]).
In the molecular hydrogen case, the molecular fraction increases with
column density due to the drop in FUV field and larger ${\rm H_2}$ 
self-shielding.
However, in the case of 
the $\cicol/\ciicol$ ratio, it is just the dust extinction that leads
to the increase in the ratio with column.


\subsection{$\chi^2$ Results}
\label{subsec:chisquareresults}

\begin{figure}
\plotone{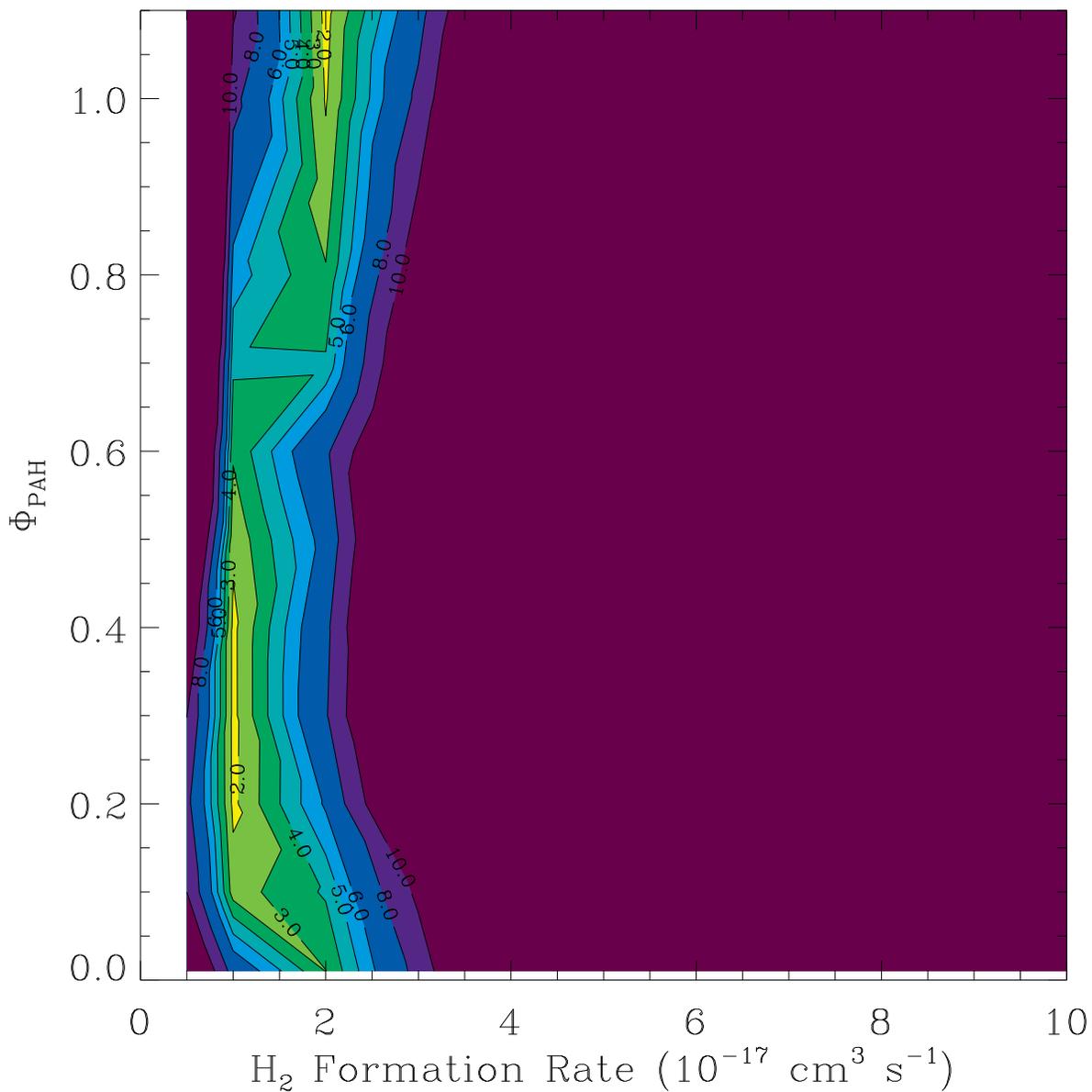}
\caption{$\chi^2$ plot of $\Phi_{\rm PAH}$ versus ${\rm H_2}$
formation rate $R$. Contour levels are 
$\chi^2 = 2$, 3, 4, 5, 6, 8, and 10. 
Observations are restricted to cloud column densities
in the range $0.03 \la A_{\rm V} \la 0.25$ and 
$N_{\rm H_2} < 10^{17}$ cm$^{-2}$. 
The minimum value of $\chi^2_{\rm min} = 1.6$ with
10 sources included.
}
\label{fig:chiav025}
\end{figure}

We next present the $\chi^2$ results obtained by binning the
sources in four $A_{\rm V}$ bins. But note that for each source
model, we fix $ A_{\rm V}$ to the observed column. 
Figure \ref{fig:chiav025} shows the results for
sources in the range $0.03 \al A_{\rm V} \al 0.25$ and
$N_{\rm H_2} < 10^{17}$ cm$^{-2}$.
There appears to be two local minima: one at  
low $R$ values with 
$R = 1\times 10^{-17}$ ${\rm cm^{3}}$ ${\rm s^{-1}}$ 
and $0.2 \al \Phi_{\rm PAH} \al 0.4$ and one at
$R\approx 2\times 10^{-17}$ ${\rm cm^{3}}$ ${\rm s^{-1}}$ 
and $\Phi_{\rm PAH}\ag 1$. The $\cicol/\ciicol$ ratios
in this $A_{\rm V}$ range are $\al 10^{-3}$ and both 
equations (\ref{eq:Agnpahm}) and (\ref{eq:Agnpahn})  apply.
At low $\Phi_{\rm PAH}$ the ${\rm C^+}$ destruction by 
electron recombinations dominate (right hand side of eq.\ 
\ref{eq:Agnpahn})
and the dependence on 
$\Phi_{\rm PAH}$ drops out. Moving to the upper right in
the $\chi^2$ plot corresponds to lower densities and higher
temperatures. There, electron recombination is less effective
in producing \ion{C}{1}, the \ion{C}{1} abundance drops, 
and the fits become poorer. At sufficiently
high $\Phi_{\rm PAH}$, the PAH reactions dominate, the
\ion{C}{1} abundance rises, and the fits improve. We shall demonstrate
that the low $R$ and $\Phi_{\rm PAH}$ minimum corresponds to 
gas densities that are closer to those expected for the diffuse
ISM and are therefore favored. 
The minimum $\chi^2_{\rm min} = 1.6$ and lies within a closed set of contours.

\begin{figure}
\plotone{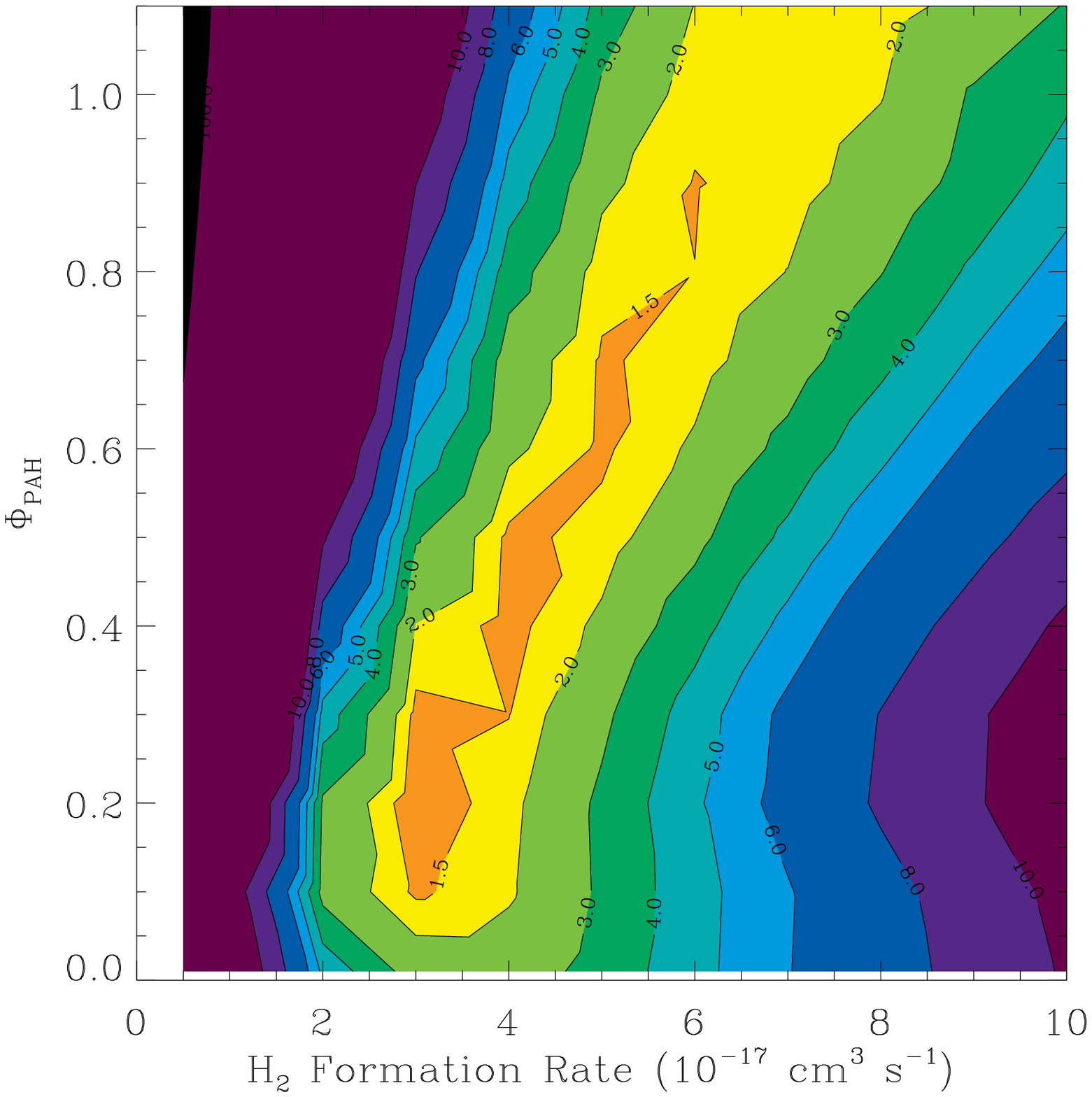}
\caption{$\chi^2$ plot of $\Phi_{\rm PAH}$ versus ${\rm H_2}$
formation rate $R$. Contour levels are 
$\chi^2 = 1.5$, 2, 3, 4, 5, 6, 8, and 10. 
Observations are restricted to cloud column densities
in the range $0.25 \la A_{\rm V} \la 0.75$ and 
$N_{\rm H_2} > 10^{18}$ cm$^{-2}$.
}
\label{fig:chiav075}
\end{figure}

\begin{figure}
\plotone{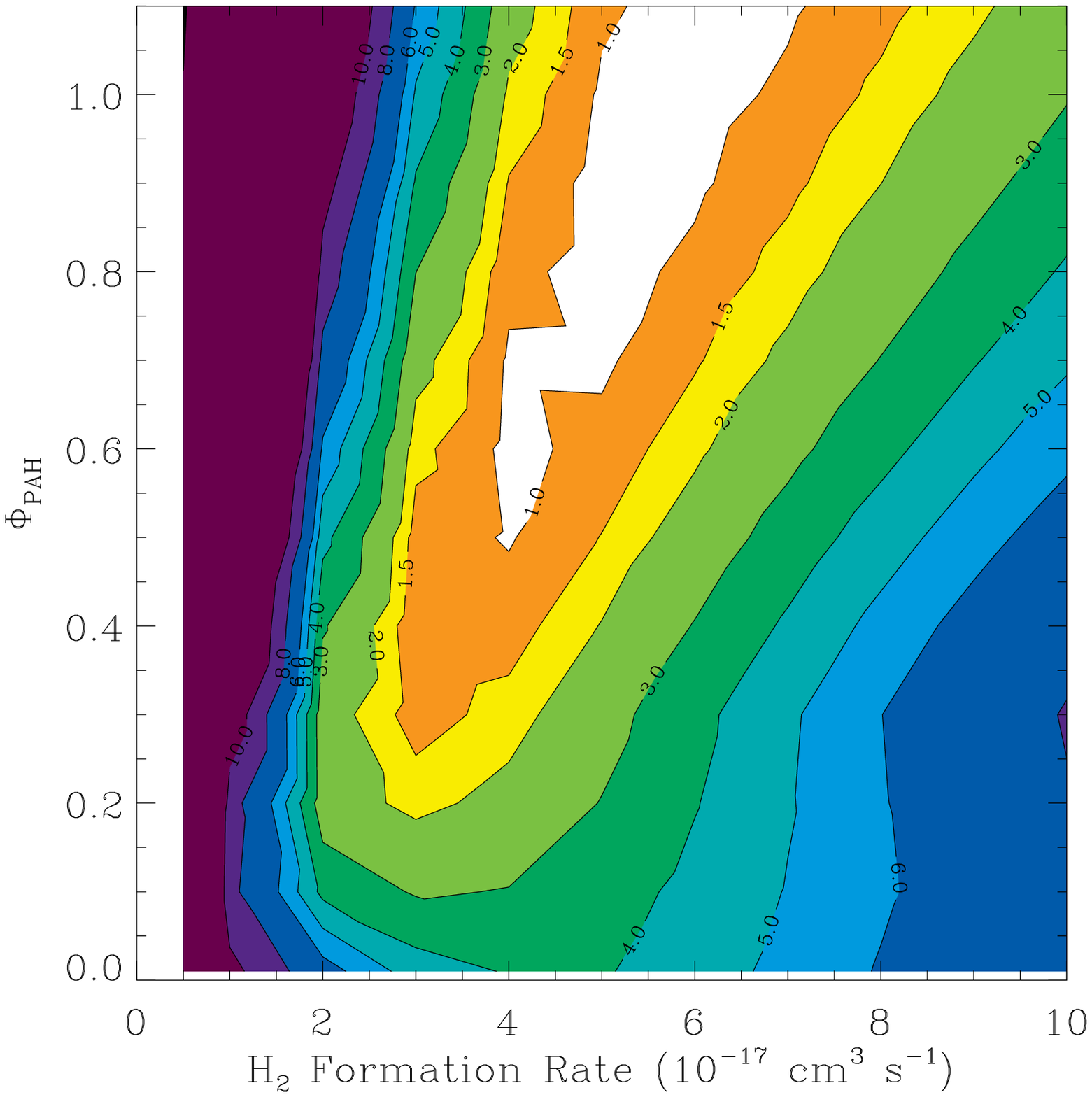}
\caption{$\chi^2$ plot of $\Phi_{\rm PAH}$ versus ${\rm H_2}$
formation rate $R$. Contour levels are 
$\chi^2 = 1$, 1.5, 2, 3, 4, 5, 6, 8, and 10. 
Observations are restricted to cloud column densities
in the range $0.75 \la A_{\rm V} \la 1.25$ and
$N_{\rm H_2} > 10^{18}$ cm$^{-2}$. 
}
\label{fig:chiav125}
\end{figure}

\begin{figure}
\plotone{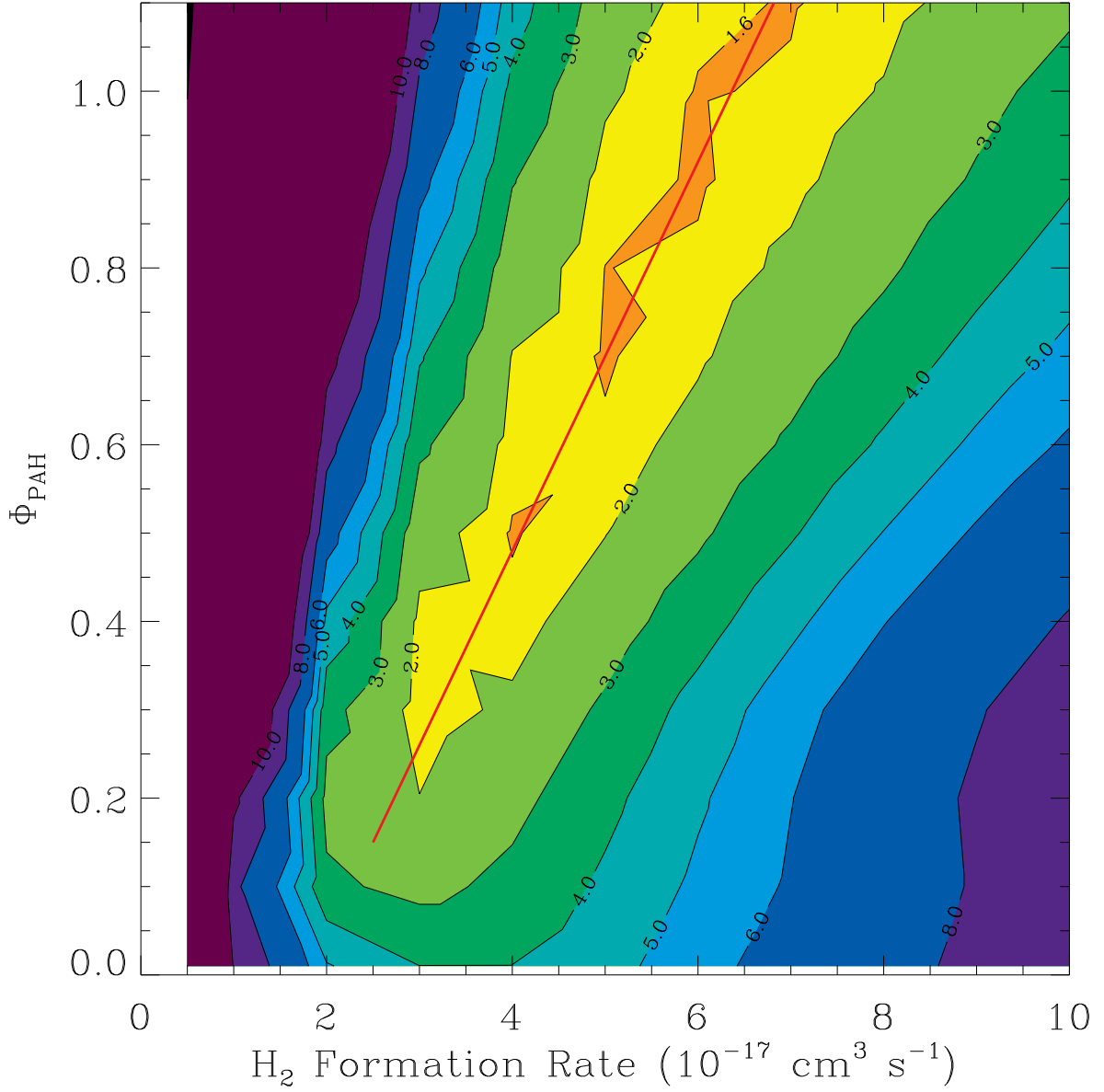}
\caption{$\chi^2$ plot of $\Phi_{\rm PAH}$ versus ${\rm H_2}$
formation rate $R$. Contour levels are 
$\chi^2 = 1.6$, 2, 3, 4, 5, 6, 8, and 10. 
Observations are restricted to cloud column densities
in the range $0.25 \la A_{\rm V} \la 2.13$ and $N_{\rm H_2} > 10^{18}$
cm$^{-2}$. 
The minimum trough is given by 
$\Phi_{\rm PAH} = 0.22 (R/10^{-17}\,\, 
{\rm cm^{3}}\,\, {\rm s^{-1}}) -0.40$
}
\label{fig:chiav025213noeps}
\end{figure}

We next show the results for
sources in the remaining $A_{\rm V}$ ranges
$0.25 \al A_{\rm V} \al 0.75$
(Fig.\ [\ref{fig:chiav075}]:  10 sources), 
$0.75 \al A_{\rm V} \al 1.25$
(Fig.\ [\ref{fig:chiav125}]:  12 sources), and 
$0.25 \al A_{\rm V} \al 2.13$
(Fig.\ [\ref{fig:chiav025213noeps}]: 28 sources). 
In these $\chi^2$
plots we do not consider the exact position of 
the minimum to be as 
significant as in Figure \ref{fig:chiav025} since it
generally lies in a trough of low values within 
open contours. 
An approximate fit to the minimum trough is given by
\beq
   \Phi_{\rm PAH} = 0.22 \left (\frac{R}{10^{-17}\, 
              {\rm cm^{3}}\,\, {\rm s^{-1}}} \right ) -0.40~~~R\ag 
      3\times 10^{-17} ~{\rm cm^{3}}\,\, {\rm s^{-1}}~~~.
\label{eq:phitrough}
\eeq
and shown in Figure \ref{fig:chiav025213noeps}. 



The $\chi^2$ contours 
shows that $R$ and $\Phi_{\rm PAH}$ are correlated.
The minimum
lies at roughly a 45 degree angle with $\Phi_{\rm PAH}$
proportional to $R$. 
For a given observation,
that is a fixed $f_{\rm H_2}$, and $A_{\rm V}$, 
from equation 
(\ref{eq:Afh2highcol}), the
ratio $Rn/G_0$ must be constant along the line of
minimum $\chi^2$. In addition, for a fixed $\cicol/\ciicol$,
from equations (\ref{eq:Agnpahm}) and (\ref{eq:Agnpahn}) 
the ratio $G_0/(n\Phi_{\rm PAH})$ must be constant along
the line of minimum $\chi^2$. Together these imply that
$R/\Phi_{\rm PAH}$ must be constant or $R\propto \Phi_{\rm PAH}$.

There is a suggestion from  our figures that the
best fit $\Phi_{\rm PAH}$ and $R$ increases
with column density. For  
$0.25 \al A_{\rm V} \al 0.75$
(Fig.\ [\ref{fig:chiav075}]) the minimum $\chi^2$ lies at
$R= 3\times 10^{-17}$ ${\rm cm^3}$ ${\rm s^{-1}}$, 
and $\Phi_{\rm PAH}=0.2$,
while for 
$0.75 \al A_{\rm V} \al 1.25$
(Fig.\ [\ref{fig:chiav125}]), there is a minimum trough that extends
from $R= 4\times 10^{-17}$ ${\rm cm^3}$ ${\rm s^{-1}}$, 
and $\Phi_{\rm PAH}=0.4$ to 
$R= 6\times 10^{-17}$ ${\rm cm^3}$ ${\rm s^{-1}}$, 
and $\Phi_{\rm PAH}=1.0$. 
If this effect is real then it is not entirely clear what
is the physical reason behind it. Perhaps shock processing
along low $A_{\rm V}$ lines of sight tend to reduce grain surface area
and produce smaller $R$ (and therefore smaller $\Phi_{\rm PAH}$ since
$\Phi_{\rm PAH} \propto R$). The statistics however, are rather poor
for separate $A_{\rm V}$ bins and additional observations are required
to confirm the trend.

\begin{figure}
\plotone{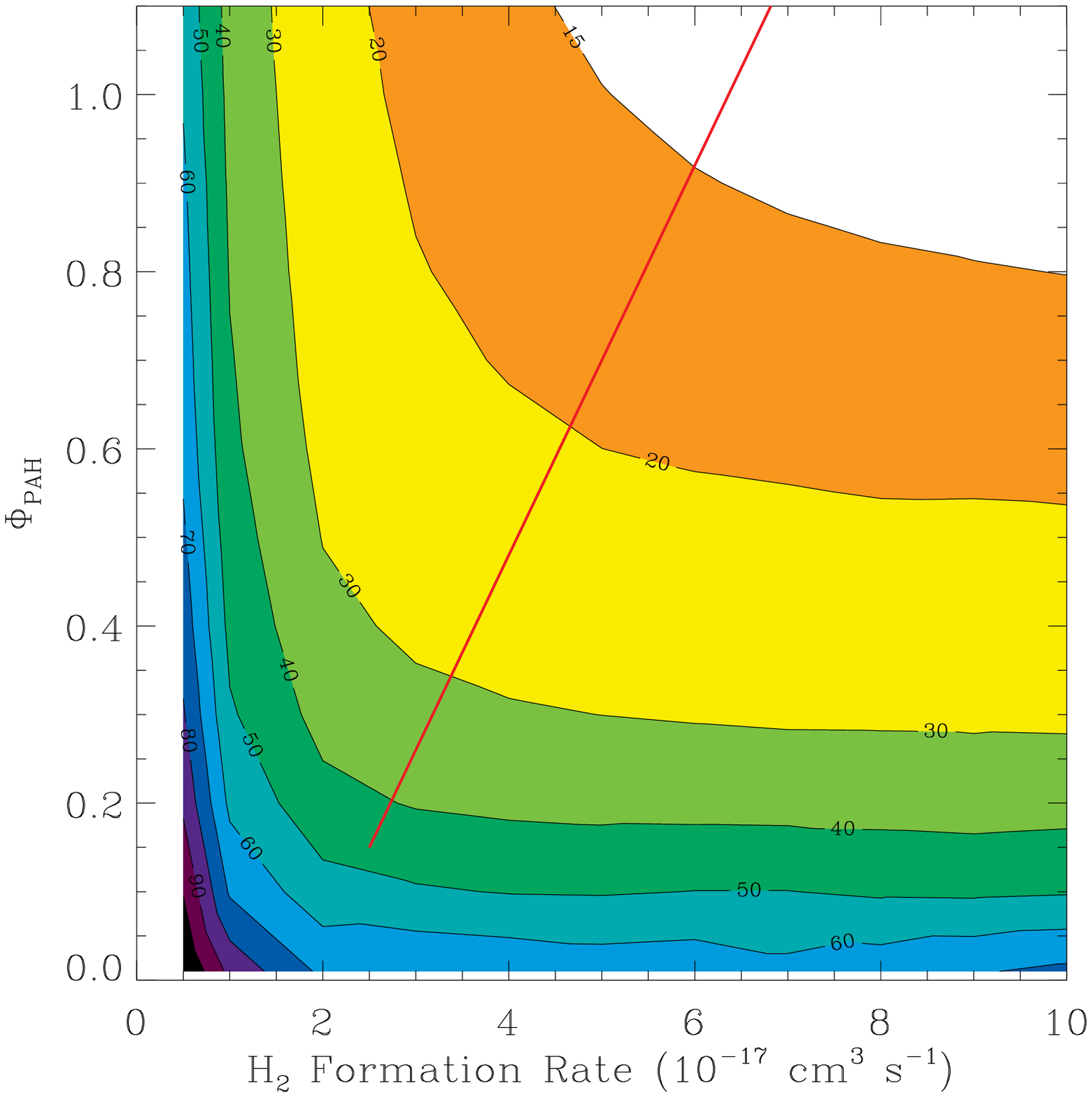}
\caption{Mean density plot of $\Phi_{\rm PAH}$ versus ${\rm H_2}$
formation rate $R$. Contour levels are 
$< n > = 15$, 20, 30, 40, 50, 60, 70, 80, 90, and 100
${\rm cm^{-3}}$. 
Observations are restricted to cloud column densities
in the range $0.25 \la A_{\rm V} \la 2.13$ and $N_{\rm H_2}>10^{18}$. 
The minimum trough from 
Fig.\ (\ref{fig:chiav025213noeps})
is given by 
$\Phi_{\rm PAH} = 0.22 (R/10^{-17}\,\, 
{\rm cm^{3}}\,\, {\rm s^{-1}}) -0.40$
}
\label{fig:denav025213noeps}
\end{figure}

Note that as the ${\rm H_2}$ formation rate $R$ increases
the density $n$ must decrease (for fixed $G_0$) in
order to maintain the same molecular fraction.
At some large value of $R$ the 
densities become much
lower than typically observed in the diffuse ISM. 
In Figures
\ref{fig:chiav025} and
\ref{fig:chiav025213noeps} 
each $R$ and $\Phi_{\rm PAH}$ grid point
has several sources contributing to the $\chi^2$ value 
and there is a best fit density for each source. 
We show in Figures \ref{fig:denav025cp} and 
\ref{fig:denav025213noeps} 
the average density $\langle n \rangle$, 
obtained by averaging the best fit density 
over the sources contributing 
to each $R$ and $\Phi_{\rm PAH}$ grid point.
Recall that in computing the $\chi^2$ grid we are
finding the best fit ratio $G_0/n$ and have fixed
$G_0$ at 1.7. Thus the density could be higher
for higher $G_0$. 
Also shown in Figure \ref{fig:denav025213noeps}
is the 
line along the $\chi^2$ minimum trough 
(eq.[\ref{eq:phitrough}]).
The density plot (Fig.\ \ref{fig:denav025213noeps}) combined with
the $\chi^2$ plot (Fig.\ \ref{fig:chiav025213noeps}) shows that the  
typical diffuse cloud ($n \sim 30$ ${\rm cm^{-3}}$, $G_0\sim 1.7$) 
lies at $R \approx 3.5\times 10^{-17}$ ${\rm cm^{3}}$ ${\rm s^{-1}}$
and $\Phi_{\rm PAH} = 0.4$. For the low column densities, the contours
centered on  $R
\approx 1\times 10^{-17}$  ${\rm cm^{3}}$ ${\rm s^{-1}}$, and
$\Phi_{\rm PAH} = 0.2-0.4$
correspond to gas densities closer to that expected in diffuse
gas and we therefore favor these over the larger $R$ and
$\Phi_{\rm PAH}$ values.


\subsection{Effects of ${\rm C^+}$ Recombination Rate}
\label{subsec:dielectric}
\begin{figure}
\plotone{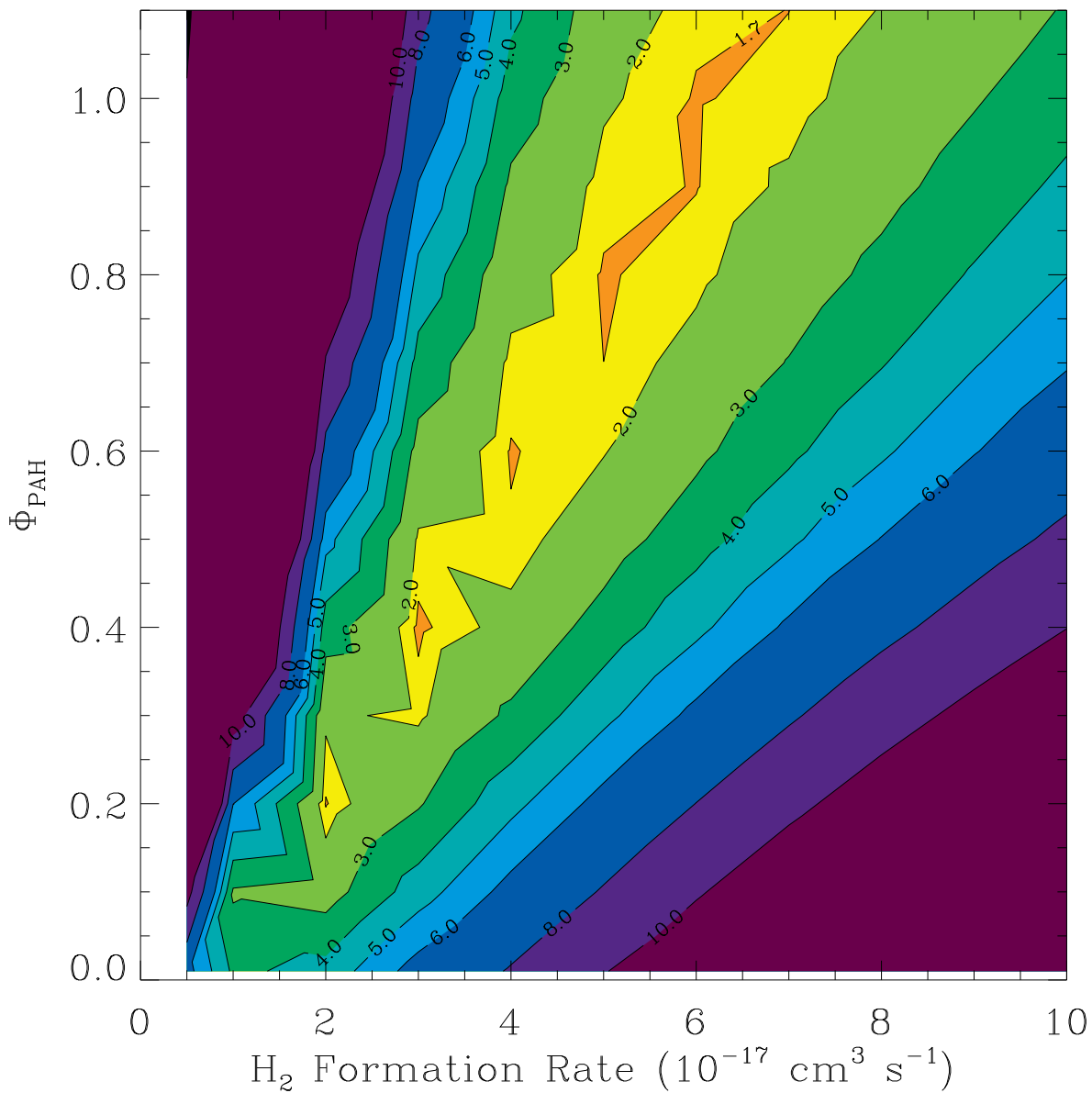}
\caption{$\chi^2$ plot of $\Phi_{\rm PAH}$ versus ${\rm H_2}$
formation rate $R$ using ${\rm C^+}$ recombination 
rate from \cite{nahar1997}.
Contour levels are 
$\chi^2 = 1.7$, 2, 3, 4, 5, 6, 8, and 10. 
Observations are restricted to cloud column densities
in the range $0.25 \la A_{\rm V} \la 2.13$ and $N_{\rm H_2} > 10^{18}$
cm$^{-2}$. 
}
\label{fig:chiav025213noepsnocp}
\end{figure}

The  recombination rate of ${\rm C^+}$ with electrons  
is most important at low $\phi_{\rm PAH}$ or high 
values of $G_0/n$ (see Appendix~\ref{appen:estimatinggn}).
The total rates (including dielectronic recombination) 
of \cite{nahar1997} are higher by a factor 
of $\sim 2$ at 100 K compared to  
\cite{altun2004}
and further diverge at lower temperatures.   
The $\cicol/\ciicol$ ratio can thus vary by a factor of
 $\sim 2$  when ${\rm C^+}$ recombination
with electrons dominates the production of \ion{C}{1}.
In previous sections of this paper we have used the
higher rates, but since the rates are uncertain, we have also
investigated the effects of the lower rates. Figure 
\ref{fig:chiav025213noepsnocp} shows the resulting 
$\chi^2$ plot for the range of column densities
$0.25 \la A_{\rm V} \la 2.13$ and $N_{\rm H_2} > 10^{18}$ cm$^{-2}$. 
The effects of the \cite{nahar1997}
rates more narrowly constrain the best values of $\phi_{\rm PAH}$ 
and $R$ by excluding the parameter space of low $\phi_{\rm PAH}$ and
large $R$. This is because  of the decreased importance of the 
electron recombination rates and thus the observed $\cicol/\ciicol$ ratio can 
only be achieved through higher PAH reaction rates. The best fit
minimum trough, however, is not changed significantly compared to 
Figure \ref{fig:chiav025213noeps},
with  only a slight
extension to lower $R \sim 2.0\times 10^{-17}$ $\rm cm^{3}$ ${\rm s^{-1}}$
and $\phi_{\rm PAH}\sim 0.2$.

\subsection{$A_{\rm V} < 0.25$ Lines of Sight}
\label{subsec:lowav}

\begin{figure}
\plotone{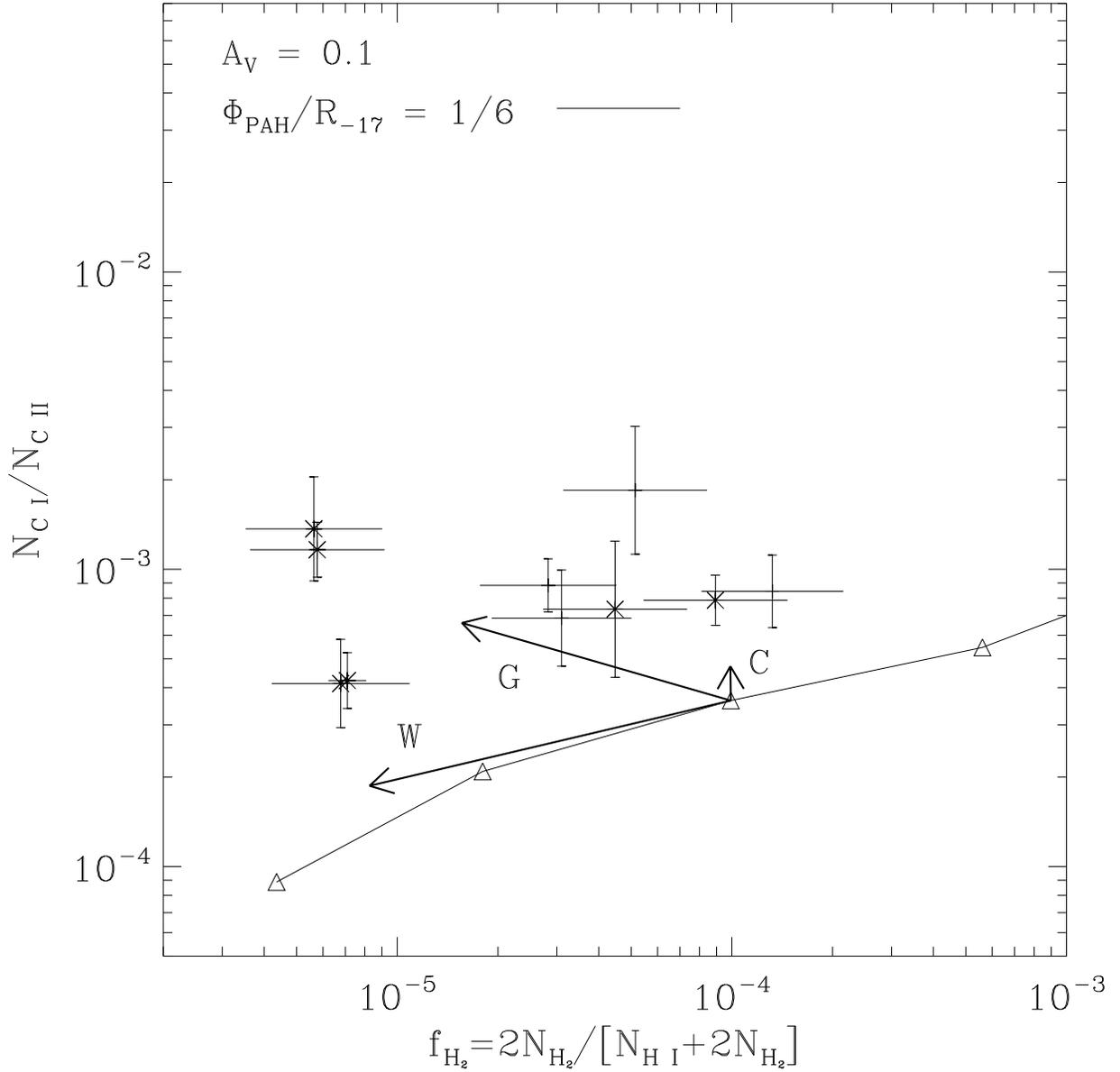}
\caption{
Column density ratio $\cicol/\ciicol$ versus molecular 
hydrogen fraction $f_{\rm H_2} = 2N_{\rm H_2}/[\hicol + 2N_{\rm
  H_2}]$. Observations are shown for cloud column densities
in the range $0.03 \la A_{\rm V} \la 0.25$ and $N_{\rm H_2} < 10^{17}$
cm$^{-2}$.
Clouds with $A_{\rm V} < 0.1$ are indicated with an 
``$\times$''. 
Curves show model results for cloud column $A_{\rm V} =
0.1$, and
$\Phi_{\rm PAH}=0.5$, 
$R = 3\times 10^{-17}$ ${\rm cm^{3}}$ ${\rm s^{-1}}$
($\Phi_{\rm PAH}/R = 1/6$; {\em solid}).
The ratio $G_0/n$ varies along each model curve 
with higher 
$G_0/n$ yielding smaller values of $f_{\rm H_2}$.   
Individual models are shown with $n = 10$, 20, 30, and 40
${\rm cm^{-3}}$
and $G_0 = 5.1$ ($\triangle$) .
Arrows 
show the affects of 
lines of sight which graze a larger ($A_{\rm V}=1$) cloud ('G'),
have cosmic rays enhanced by a factor of 10 ('C'), and have half
the column density in the WNM phase ('W'). 
}
\label{fig:ratioav025var}
\end{figure}

For our low column density lines of sight, the combination
of a low molecular hydrogen fraction, $f_{\rm H_2}$ and a relatively
high $\cicol/\ciicol$ ratio indicate that $R$ must be low or at
least a low ratio of $R/\Phi_{\rm PAH}$ (ref.\ Figs.\ [\ref{fig:ratioav025}] 
and [\ref{fig:chiav025}]).
Are there unusual physical conditions, or additional
destruction processes that
are acting in these regions which could  account for
the low ${\rm H_2}$ columns? 
 We first consider the effects of
warm neutral medium (WNM) along the line of sight. The WNM consists
of low density ($n \sim 0.3 $ ${\rm cm^{-3}}$) warm ($T\sim 8000$ K) gas.
Such gas will likely contain 
\ion{C}{2} and \ion{H}{1}, but very little \ion{C}{1} and ${\rm H_2}$. 
This is because the formation rates of \ion{C}{1} and ${\rm H_2}$ are
proportional to $n^2$, while the destruction rates are proportional
to $n$ and thus the abundance drops in lower density gas. In addition, at
high temperatures the ${\rm C^+}$ recombination rates fall. The ${\rm H_2}$ formation rate
may also be lower in the WNM because at high temperatures the \ion{H}{1} atoms 
colliding with grains will more likely bounce than stick \citep{burke1983}. 
We tested results for an $A_{\rm V} = 0.1$ cloud
with $G_0 = 5.1$ and $n = 30$  ${\rm cm^{-3}}$ in which
half the column is WNM gas. Here we have included the density
and temperature effects in the rates
but not reduced $R$ due to 
a smaller sticking coefficient 
(see Fig.\ \ref{fig:ratioav025var} arrow `W'). The model 
point which includes WNM moves towards lower $\cicol/\ciicol$ ratio 
(by a factor of $\sim 2$) and lower
$f_{\rm H_2}$ (by a factor of $\sim 10$). Part of the
drop in $f_{\rm H_2}$ is due to lower ${\rm H_2}$ 
self-shielding in the $n=30$ ${\rm cm^{-3}}$ cloud. 
The resulting vector moves  
away from the low $A_{\rm V}$ observations, and along
a path which is nearly parallel to the standard models.
Because the addition of WNM along the line-of-sight only
moves points parallel to the standard models, it is unlikely
that the presence of a WNM component can explain the ${f_{\rm H_2}}$
and $\cicol/\ciicol$ results at low column.

We next consider the effects of a higher cosmic ray ionization rate along
low column density lines of sight. 
The effects of increasing the primary ionization rate by a factor of 10
has little affect on the ${\rm H_2}$ columns since
photodissociation dominates the destruction of ${\rm H_2}$.
 The $\cicol/\ciicol$ ratio increases slightly (by $\sim 30$\%)
due to the greater electron abundance from cosmic ray ionizations
(Fig.\ \ref{fig:ratioav025var} arrow `C').
We conclude that neither WNM nor enhanced cosmic ray rates 
can explain the low $A_{\rm V}$ observations. 

Another possibility is that the low $A_{\rm V}$ lines of sight
are strongly affected by grain processing in interstellar shocks.
\cite{jones1996} demonstrated that sputtering and
grain-grain collisions can vaporize grains, and reduce the grain
surface area in sufficiently fast 
$v_{\rm s} \al 150$ km ${\rm s^{-1}}$ shocks. 
In the interstellar
medium, shock speeds tend to scale with $n^{-1/2}$, so that lower
density gas will experience higher velocity shocks. Low $A_{\rm V}$
gas may have higher fractions of low density WNM gas. We note however,
that shocks of speeds $v_{\rm s} \approx 100$ km ${\rm s^{-1}}$ tend
to shatter the larger grains and produce a large population of small
grains which increases the surface area. We speculate that 
another  type of grain processing
occurring in shocks is to sputter the surfaces ``clean'' and thereby
modify the characteristics of the atomic hydrogen adsorption 
sites (e.g., binding energy).

A final possibility is that the low $A_{\rm V}$ columns are lines
of sight which graze the edges of larger clouds. In this case, the
typical shielding column density (which is weighted towards
the smallest column direction, i.e., perpendicular to the line-of-sight)
is much less than the line-of-sight column that we measure.
We examine this 
possibility with a simple test. We take a cloud with a total
$A_{\rm V}=1$ through the center and consider a ray which passes
through the cloud with a column of $A_{\rm V} = 0.1$. At the midpoint,
such a  ray
passes through a minimum depth of $A_{\rm V} = 2.5\times 10^{-3}$ measured
from the cloud surface. We take the $n_{\rm H_2}$, $n_{\rm H\, I}$, 
$n_{\rm C\, I}$, and $n_{\rm C\, II}$ abundances as a function of 
cloud radius from model runs for $A_{\rm V} = 1$, $G_0 = 5.1$, 
and $n = 30$ ${\rm cm^{-3}}$. We then integrate the 
$\hicol $, $N_{\rm H_2}$, $\cicol$, and $\ciicol$  column densities
along the grazing line of sight. Results are shown in Figure 
\ref{fig:ratioav025var}
as an arrow labeled `G'. We see that compared to an $A_{\rm V} = 0.1$ cloud, 
the $f_{\rm H_2}$ decreases due
to the diminished ${\rm H_2}$ self-shielding at the cloud edge. 
Only a small molecular column exists between the FUV illuminated 
surface and points along the ray and thus the self-shielding is
ineffective and the ${\rm H_2}$ is dissociated.
We also see that the $\cicol/\ciicol$ ratio
rises. This is because the FUV field is lower by a factor of two compared
to the $A_{\rm V} = 0.1$ cloud since the radiation can not penetrate
from the backside of the larger cloud. The lower field reduces the 
photoionization rate
of C to ${\rm C^+}$ and increases the abundance of ${\rm PAH^-}$ 
which converts ${\rm C^+}$ to C. (The lower field also reduces the 
${\rm H_2}$ dissociation rate but this reduction is offset by the
lack of ${\rm H_2}$ self-shielding). The results significantly
move the model points towards the observations. We consider
grazing incident lines of sight 
a potential explanation for the low column density observations
without resorting to unusual grain properties.  

For $A_{\rm V} \al 0.25$ we find two lines of sight to 
be anomalous. These are
$\epsilon$ Per ($A_{\rm V} = 0.16$, $f_{\rm H_2} = 0.20$) 
and 59 Cyg  ($A_{\rm V} = 0.11$, $f_{\rm H_2} = 0.19$).
These lines of sight have molecular fractions which
are $\ag 10^3$ times greater than others  with similar (low) 
column densities and require different rates or environments
than the low $f_{\rm H_2}$ cases. 
Including $\epsilon$ Per 
and 59 Cyg in the $\chi^2$ tests for the 
$0.03 \al A_{\rm V} \al 0.25$, $N_{\rm H_2} < 10^{17}$ sources
raises the minimum $\chi^2$ from 1.6 to 9.2 and thus these sources
can well be considered to be part of a separate population than the
rest of the low $A_{\rm V}$ and low $f_{\rm H_2}$
lines of sight.
The $\kappa$ Ori line of sight at slightly higher
column density ($A_{\rm V} = 0.17$)  
is part of the low $f_{\rm H_2}$ group as are 
 $\epsilon$ Ori
$(A_{\rm V} = 0.14)$ and 15 Mon $(A_{\rm V} = 0.13)$.
One explanation for these peculiar sight lines could be
that shocks of speeds $v_s\approx 100$ km ${\rm s^{-1}}$
have significantly increased the population of small
grains. The resulting increase in grain surface area
will increase the FUV extinction (and decrease the 
${\rm H_2}$ photodissociation rate) and also increase the
${\rm H_2}$ formation rate. We find that with the FUV extinction and
${\rm H_2}$ formation rate a factor of three higher than normal 
we can reproduce the $\cicol/\ciicol$ ratio and molecular 
fraction for the $\epsilon$ Per line of sight. 
Unfortunately, there are
no low resolution IUE data for $\epsilon$ Per to examine
the FUV rise in the extinction curve. We also note that 
59 Cyg is a variable Be star and column densities for this
line of sight are highly uncertain (D.\ Massa, private
communication).



\subsection{$A_{\rm V} > 0.25$ Lines of Sight}
\label{subsec:highav}

At columns densities $A_{\rm V} > 0.25$ we found two points
to be anomalous: 23 Ori and $\pi$ Sco. \cite{weingartner2001}
modeled the line of sight towards 23 Ori 
and found that they
had some difficulty matching the high $\cicol$ column even when 
recombination on grains was included. They proposed several 
additional processes
which might be working to enhance the C abundance including the
 dissociative recombination of ${\rm CH^+}$. The ${\rm CH^+}$ abundance
can be enhanced compared to normal equilibrium chemistry due to 
the effects of non-thermal chemistry 
\citep{flower1998,joulain1998,zsargo2003}. If  
turbulent velocities are sufficiently large ($v\ag 3-4$ km ${\rm s^{-1}}$)
and turbulent dissipation can drive reactions with temperature barriers, then
a large column of ${\rm CH^+}$ can be produced \citep{joulain1998}. 
 
Another process which might account for the anomalous columns is 
the time dependence of the chemistry. The line of sight towards
23 Ori shows a large $\cicol/\ciicol$ ratio at small molecular
fractions $f_{\rm H_2}$ compared to the equilibrium abundances. 
Consider a parcel of gas which has been recently shocked (i.e., 
${\rm H_2}$ dissociated and \ion{C}{1} ionized). The
molecular hydrogen is produced at a rate of $\sim 3\times 10^{-17}n^2$
${\rm s^{-1}}$. The time to double the current fractional abundance 
$n_{\rm H_2}/n \sim 5\times 10^{-3}$ is about 
$t\sim 5\times 10^{-3}/(3\times 10^{-17}n)$ s or $t\sim 5.3\times 10^{4}$
yr for $n=100$ ${\rm cm^{-3}}$. On the other hand, 
the time to double the current ratio $\cicol/\ciicol\sim 0.01$ 
is about $t\sim 0.01/(\Phi_{\rm PAH}2.4\times 10^{-7}n_{\rm PAH^-})$ s
or $t\sim 62$ yr for $n=100$ ${\rm cm^{-3}}$, where we have used 
${\rm C^+}$ recombination with ${\rm PAH^-}$,
$\Phi_{\rm PAH} = 0.5$, and
$n_{\rm PAH^-}$ from equation (\ref{eq:Anpahm}).
The time to reach the equilibrium $\cicol/\ciicol$ ratio is much shorter than 
the time to reach the equilibrium $f_{\rm H_2}$ fraction. Thus, for 
a recent ($t\al 5\times 10^{4}$ yr) passage of a shock, the 
$\cicol/\ciicol$  ratio has had time to reach equilibrium while 
the ${\rm H_2}$ fraction is dissociated compared to equilibrium.

For $\pi$ Sco consider a parcel of molecular gas which has been recently
illuminated by a source of FUV radiation. The time to dissociate the current
$f_{\rm H_2}\sim 0.01$ at cloud center is 
$t\sim (G_0 I \beta[N_{\rm H_2}] \exp[-2.5 A_{\rm V}])^{-1}$
or $t \sim 1.1\times 10^{6}$ yr, where we have 
used $G_0=5.1$, $N_{\rm H_2} = 1\times 10^{19}$
${\rm cm^{-2}}$, $\beta[1\times 10^{19}\,{\rm cm^{-2}}]=1.7\times 10^{-4}$
(Eq.[\ref{eq:dbself}]), and $A_{\rm V} = 0.13$.
The time to ionize the current $\cicol/\ciicol \sim10^{-4}$ ratio is
about $t\sim (2.1\times 10^{-10}G_0 \exp[-2.6 A_{\rm V}])^{-1}$ s
or $t\sim 41$ yr. Thus for a molecular clump illuminated
within the past $\sim 10^6$ yr, the $\cicol/\ciicol$ ratio is in equilibrium
while the ${\rm H_2}$ fraction is much higher than its equilibrium 
value.

One complicating factor which we have not yet considered is the 
effects of multiple clump components along the line of sight. 
In general, taking the observed column density and splitting 
it up into multiple clouds exposes more surfaces to the 
interstellar radiation field. In the case of optically
thick clumps, 
this will decrease the 
$\cicol/\ciicol$ ratio and decrease the molecular fraction 
$f_{\rm H_2}$. 
For the case of optically thin clumps, the 
$\cicol/\ciicol$ ratio will not change but 
$f_{\rm H_2}$ will drop due to reduced self-shielding.
However, the effects of multiple clumps is important only for
components of approximately equal column density. If one component
dominates then the remaining smaller components do not contribute
much to the total column. Since in general, the UV absorption
spectroscopy observations are dominated
by a single component, our neglect of multiple clouds
should not significantly alter our results.



\section{Summary}
\label{sec:summary}

We have carried out an analysis of observations
of the \ion{C}{1}, \ion{C}{2}, \ion{H}{1}, and 
${\rm H_2}$ column densities in the diffuse ISM
towards 42 lines of sight spanning a range of
column densities $0.03 \le A_{\rm V} \le 2.13$.
We have fitted these columns using a 2-sided
PDR model which simultaneously solves for the
thermal and chemical balance throughout the layer.
The relative columns of \ion{H}{1} and ${\rm H_2}$ are dependent 
on the rate coefficient $R$ for ${\rm H_2}$  formation on grain surfaces, 
on the ratio $G_0/n$,
and on the total column $N$ (or $A_{\rm V}$) of the illuminated cloud.  The
relative columns of \ion{C}{1}  and \ion{C}{2} depend on the electron 
and PAH abundances, the rate coefficients for neutralization of C$^+$ with 
PAHs (or $\Phi _{\rm PAH}$), the total column in the illuminated cloud, and $G_0/n$.
Observations set the total columns of each line of sight plus the
fractional abundances of H, ${\rm H_2}$, C, and ${\rm C^+}$.  Our PDR models self
consistently determine the electron, PAH$^+$, PAH, and PAH$^-$ abundances
for each model.  We have performed a $\chi^2$ test using $R$,
$\Phi _{\rm PAH}$, and $G_0/n$ as our free parameters to find the best
fits to the fractional abundances along each line of sight.

\begin{figure}
\plotone{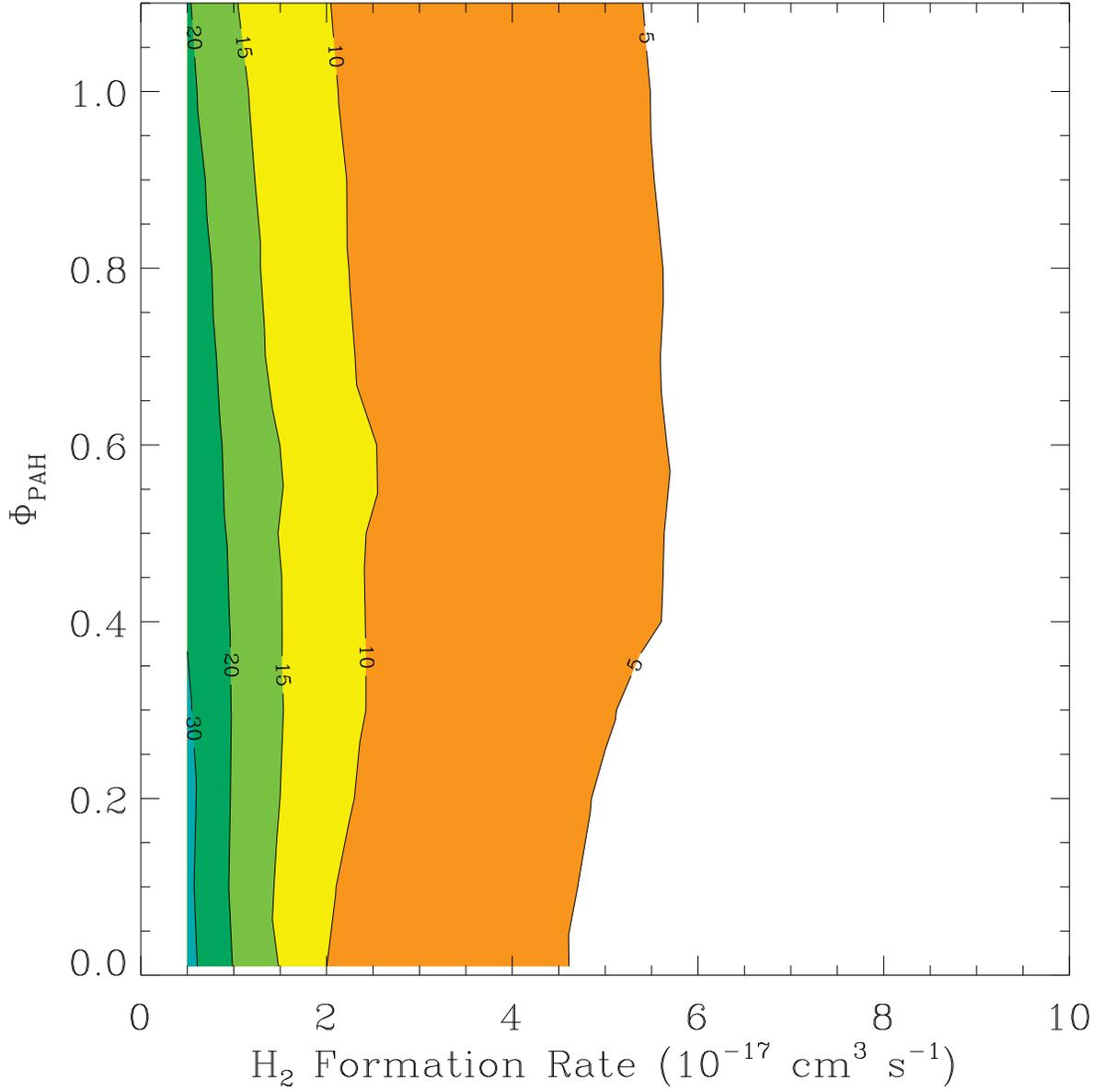}
\caption{Mean density plot of $\Phi_{\rm PAH}$ versus ${\rm H_2}$
formation rate $R$. Contour levels are 
$< n > = 5$, 10, 15, 20, and 30
${\rm cm^{-3}}$. 
Observations are restricted to cloud column densities
in the range $0.03 \la A_{\rm V} \la 0.25$ and $N_{\rm H_2} < 10^{17}$
cm$^{-2}$. 
}
\label{fig:denav025cp}
\end{figure}

The results are shown in Figures \ref{fig:chiav025} through \ref{fig:denav025213noeps}, 
where we show the $\chi ^2$
contours in the $R$, $\Phi _{\rm PAH}$ plane, given the best fitting $G_0/n$
for each individual line of sight.  Note that Figures 
\ref{fig:chiav025} through \ref{fig:denav025213noeps}
 are
the $\chi ^2$ for these values of $R$, $\Phi _{\rm PAH}$ for all 42 sources.
Our main conclusions are as follows:

\noindent 
1. For low column ($A_{\rm V} < 0.25$) lines of sight, and for the low
molecular columns and fractions that accompany them 
($N_{\rm H_2}< 10^{17}$ cm$^{-2}$, $f_{H_2} < 10^{-4}$), we
find $R \simeq 1 \times 10^{-17}$ cm$^3$ s$^{-1}$ and $\Phi_{\rm PAH}
\sim 0.2 - 0.4$.

\noindent
2. At higher column densities ($0.25 < A_{\rm V} < 2.13$), 
 we find $R \simeq 3.5 \times 10^{-17}$ cm$^3$ s$^{-1}$ and 
$\Phi _{\rm PAH} \sim 0.4$. 

\noindent
3. Our $\chi ^2$ fits show a correlated range of $\Phi _{\rm PAH}$ and $R$
values that provide as good fits as our ``best fits" quoted above.  For
$A_V > 0.25$, we find good fits for $\Phi _{PAH} = 0.22 (R/10^{-17} {\rm
cm}^3\ {\rm s}^{-1}) - 0.40$ as long as $R> 3 \times 10^{-17}$
cm$^3$ s$^{-1}$.  Very high values of $R$ (or therefore $\Phi_{\rm PAH}$),
however, can only be fit with very low average densities in
the clouds along the 42 sight lines.  Figure \ref{fig:denav025213noeps}
shows that for
densities (equivalently thermal pressures) thought to be typical
of local diffuse interstellar clouds, a value of $R \sim 3.5 \times 10^{-17}$
cm$^3$ s$^{-1}$ (and therefore $\Phi _{\rm PAH} \sim 0.4$)
is preferred.  This value is consistent with previous determinations
of $R$ made by comparing observations of ${\rm H_2}$ with model 
calculations \citep[e.g.,][]{jura1975}.

\noindent
4. The low values of $R$ for low column density sight lines may indicate
shock processing that decreases grain surface area or modifies 
grain surfaces along these sightlines, thereby reducing the ${\rm H_2}$ 
formation rates, or  a line of sight  which grazes a larger cloud. 


\noindent
5. Our results for $\Phi _{\rm PAH}$ indicate that the crude estimates of these
ratios adopted by \cite{wolfire2003} were roughly correct. 
They confirm the importance of PAHs in determining
the ionization level of atomic ions (in this case C vs C$^+$) in
interstellar clouds.

\acknowledgments

M.G.W. was supported in part by a NASA Long Term Space Astrophysics (LTSA) 
grant NNG05GD64G. We thank an anonymous referee for a careful reading of
our paper and for comments that improved the presentation.

\appendix

\section{Chemical Rates}
\label{appen:pahrates}

The chemical rates with PAHs were presented in \cite{wolfire2003}
Appendix C2 for reactions involving the ionization balance
of hydrogen. Here we reproduce the rates along with the Carbon rates
used in this paper. Our starting
point are the rates from \cite{draine1987} with $N_{\rm C} = 35$
carbon atoms, and disk PAHs with a radius $a=(N_{\rm C}/1.222)^{0.5}$.
We use a PAH abundance of $n_{\rm PAH} = 6\times 10^{-7}n$. 
\cite{wolfire2003} introduced the factor $\Phi_{\rm PAH}$ to 
account for a wide range of unknowns in the PAH distribution
and rate coefficients. Simple fits to the rates appropriate 
to diffuse ISM conditions 
are presented in Table \ref{tbl:reactions}.

\section{Estimating Radiation Field $G_0$ and Density $n$}
\label{appen:estimatinggn}

In this appendix we derive simple expressions to estimate the
FUV radiation field $G_0$ incident on the cloud and the cloud 
density $n$ from the observed column densities of $N({\rm H\, I})$, 
$N_{\rm H_2}$,
$\cicol$, and $\ciicol$.
In the limit of no dust extinction, the equilibrium abundance of
${\rm H_2}$ is given by 
\beq
G_0 I n_{\rm H_2} \beta [N_{\rm H_2}] = R\,n \, n_{{\rm H\, I}}~~~,  
\label{eq:Ah2eq}
\eeq
where $G_0$ is the incident FUV radiation field, $I$ the unshielded 
photodissociation rate, $\beta [N_{\rm H_2}]$ is the ${\rm H_2}$ 
self-shielding factor, and $R$ is the formation rate on grains.
For ${\rm H_2}$ columns $\al 10^{14}$ ${\rm cm^{-3}}$,
$\beta [N_{\rm H_2}]=1$ and the molecular fraction is given by
\beq
     f_{\rm H_2} = \frac{2nR}{G_0 I}\,\,\, .
\label{eq:Afh2lowcol}
\eeq

To estimate  $G_0$ and $n$ we use a 
simple expression for the ${\rm H_2}$
self-shielding formula from 
\cite{draine1996} 
\beq
      \beta [ N_{\rm H_2} ] = \left [ \frac{N_0}{N_{\rm H_2}+N_0} 
            \right ]^{3/4}~~~,
\label{eq:dbself}
\eeq
 where $N_0 = 1\times 10^{14}$ ${\rm cm^{-2}}$  is the ${\rm H_2}$ column
 where line self-shielding starts to become significant. We can integrate
equation (\ref{eq:Ah2eq}) through the cloud 
\beq 
G_0 I N_0^{3/4} \int \frac{d N_{\rm H_2}^{'}}{[N_0 + N_{\rm H_2}^{'}]^{3/4}}
  = Rn\int d N_{\rm H\, I}^{'}~~~.
\eeq
Letting $x = N_0 + N_{\rm H_2}$ the integral becomes
\beq 
G_0 I N_0^{3/4} \int_{N_0}^{N_0 + N_{\rm H_2}}
 \frac{d x^{'}}{x^{'3/4}}
  = R\, n\, N_{\rm {H\, I}}~~~.
\eeq
Carrying out the integral and solving for $G_0/n$ we have
\beq
\frac{G_0}{n} =  \frac{0.25 N_{\rm H\, I}}
              {\left \{ [N_0 + N_{\rm H_2}]^{1/4} - N_0^{1/4} \right \}
              N_0^{3/4}} \frac{R}{I}^{}~~~,
\eeq
or 
\beq
\frac{G_0}{n} =   {\cal X}_{\rm H} \frac{R}{I}^{}~~~,
\label{eq:Agn}
\eeq
with
\beq
{\cal X}_{\rm H} =  \frac{0.25 N_{\rm H\, I}}
              {\left \{ [N_0 + N_{\rm H_2}]^{1/4} - N_0^{1/4} \right \}
              N_0^{3/4}}~~~.
\label{eq:Axh}
\eeq

For $N_{\rm H_2} > N_0$ from equations (\ref{eq:Agn}), (\ref{eq:Axh}) and
\cite{hollenbach1999} the molecular fraction is given by
\beq
  f_{\rm H_2} = 2 \left ( \frac{Rn}{4G_0I} \right )^4 \left ( \frac{N}{N_0} 
              \right )^3\,\,\, .
\label{eq:Afh2highcol}
\eeq
For clouds where the optical depth of the dust becomes significant the
local FUV field within the cloud is less than the incident field. We 
assume the flux on each side of a slab is $1/2 G_0$ and the mean field, 
$\langle  G_0 \rangle$ 
within the slab is approximately at a depth of 1/4 the total column density
or at a depth of $1/4 A_{\rm V}$.  Thus the mean field is given by 
\beq
\langle G_0 \rangle = \frac {1}{2}G_0 e^{ (-\frac{3}{4} \alpha A_{\rm V})}
   + \frac {1}{2}G_0 e^{ (-\frac{1}{4} \alpha A_{\rm V})}~~~,
\eeq 
where $\alpha = 2.5$ converts the FUV extinction at visual wavelengths
to optical depth at the FUV dissociation energies of ${\rm H_2}$. Solving
for $G_0$ we have
\beq
   G_0 = \frac{2 \langle G_0 \rangle e^{\frac{2.5}{4}A_{\rm V}}}
           {1 + e^{-\frac{2.5}{2}A_{\rm V}}}~~~.
\eeq
Setting $G_0$ in equation (\ref{eq:Agn}) to the mean field 
$\langle G_0 \rangle$
we have
\beq
   \frac{G_0}{n} = \frac{2 e^{\frac{2.5}{4}A_{\rm V}}}
           {1 + e^{-\frac{2.5}{2}A_{\rm V}}} {\cal X}_{\rm H} \frac{R}{I}~~~.
\label{eq:Agmeann}
\eeq

The analytic solution is approximate since $1/4A_{\rm V}$ is an approximation
to the depth where the local field equals the mean field.
Comparing equation (\ref{eq:Agmeann}) with our numerical results 
we find we can achieve better agreement with an additional factor
of $1/2$ on the right hand side. 

A second estimate of $G_0/n$ can be obtained from the $\cicol/\ciicol$ ratio.
There are several 
destruction processes for ${\rm C^+}$ to consider depending on the 
${\rm PAH^-}$
abundance. If destruction occurs primarily through recombination on
 ${\rm PAH^-}$
then the equilibrium abundance of ${\rm C^0}$ is found by balancing 
recombination 
with photoionization of C 
\beq
   G_0 I_{\rm C} n_{\rm C^0}= \Phi_{\rm PAH} \kappa_1 n_{\rm C^+}
       n_{\rm PAH^-}~~~,
\label{eq:Arecombination}
\eeq
where $G_0I_{\rm C} = G_0 2.1\times 10^{-10}$ ${\rm s^{-1}}$ is the photoionization rate of 
${\rm C^0}$, and $\Phi_{\rm PAH} \kappa_1 = \Phi_{\rm PAH} 2.4\times 10^{-7} T_2^{-0.5}$ 
${\rm cm^3}$ ${\rm s^{-1}}$ is the ${\rm C^+}$ recombination rate with ${\rm PAH^-}$.
From \cite{wolfire2003}, the ${\rm PAH^-}$ abundance in the regime 
$n_{\rm PAH^-}/n_{\rm PAH^0} < 1$ is given by
\beq 
     n_{\rm PAH^-} = 4.8\times 10^{-5} n_e n 
               \Phi_{\rm PAH}G_0^{-1}~~~{\rm cm^{-3}},
\label{eq:Anpahm}
\eeq
 with the electron density $n_e$ and hydrogen nucleus density $n$ in units of 
${\rm cm^{-3}}$. Substituting this expression for $n_{\rm PAH^-}$ into equation
(\ref{eq:Arecombination}), and solving for $G_0/n$ we have 
\beq
\frac{G_0}{n} = 4.1 \times 10^{-3} \Phi_{\rm PAH} T_2^{-0.25}
      \left [ \frac{\cicol}{\ciicol} \right ]^{-1/2}
\label{eq:Agnpahm}
\eeq
where we have used an electron abundance of
$n_e \approx 3\times 10^{-4}n$ ${\rm cm^{-3}}$ (including metals, especially 
ionization of C,  and ionization
of H by soft X-rays), and substituted the observed ratio 
$\cicol/\ciicol$ for $n_{\rm C^0}/n_{\rm C^+}$.

If the abundance of ${\rm PAH^-}$ is low enough, then the destruction of 
${\rm C^+}$
occurs through charge exchange with ${\rm PAH^0}$ or through 
electron recombination in the
gas phase. The equilibrium abundance is then given by
\beq
G_0 I_C n_{\rm C^0}= n_{\rm C^+} (k_2 n_{\rm PAH^0} + k_3 n_e)   
\label{eq:Achargeexchange}
\eeq
with $k_2 = 8.8\times 10^{-9} \Phi_{\rm PAH}$ ${\rm cm^3\,\, s^{-1}}$,
and $k_3 = 1.8\times 10^{-11}T_2^{-0.83}$ ${\rm cm^3}\,\, s^{-1}$.
Using the ${\rm PAH^0}$ abundance of  $\sim 0.7$ times
the total PAH abundance  
($n_{\rm PAH^0}=0.7 \times 6\times 10^{-7} n$; 
Wolfire et al.\ 2003), 
and an electron abundance of $n_e \approx 3\times 10^{-4}n$ ${\rm cm^{-3}}$,
we can solve for $G_0/n$
\beq
\frac{G_0}{n} = (1.8\times 10^{-5} \Phi_{\rm PAH} + 2.6\times 10^{-5} 
            T_2^{-0.83} )    \left [\frac{\cicol}{\ciicol} \right ]^{-1}
\label{eq:Agnpahn}
\eeq
We find that the destruction of ${\rm C^+}$ by ${\rm PAH^0}$  
dominates at a limit of approximately
\beq
     \frac{G_0}{n} \ag \frac{T_2^{-1/2}\Phi_{\rm PAH}^2}
      {1.2\Phi_{\rm PAH} + 1.6 T_2^{-0.83}}
\eeq
From equation (\ref{eq:Agnpahm}), and setting $T_2=1$, and
$\Phi_{\rm PAH} = 0.5$,
this limit 
corresponds to an observed
column density ratio ${\cicol}/{\ciicol} \al 3\times 10^{-4}$. Thus for
${\cicol}/{\ciicol} \al 3\times 10^{-4}$ we use equation (\ref{eq:Agnpahn})
while for ${\cicol}/{\ciicol} >> 3\times 10^{-4}$ we use equation 
(\ref{eq:Agnpahm}). 

Similar to the $G_0/n$ estimate based on ${\rm H_2}$ columns we also
correct the incident FUV field for dust extinction within the cloud
by setting the $G_0$ in equations (\ref{eq:Agnpahm}) and (\ref{eq:Agnpahn})
to the mean FUV field $\langle G_0 \rangle$ and solve for the extinction
corrected $G_0/n$. In solving for the minimum $\chi^2$, we use the 
estimates for $G_0/n$ presented here along with a fixed value of
$G_0 = 1.7$ to provide a first guess range of $n$. If the minimum 
in $\chi^2$ is
not crossed  within the first range of $n$ we take a second range given by
$n_{\rm max} \times (n_{\rm max}/n_{\rm min})$ 
if higher densities are required
or  $n_{\rm min} \times (n_{\rm max}/n_{\rm min})^{-1}$ if lower densities 
are required
(where 
$n_{\rm max}$ and $n_{\rm min}$ refer to the maximum and
minimum densities in the range). 
The minimum in $\chi^2$ is generally found within the first or second
range. We then do a quadratic fit of the $\chi^2$ values 
as a function of density to find the density which provides the minimum
$\chi^2$ value.

\section{Variation of  $\cicol/\ciicol$ as a function of 
$G_0/n$ and $\Phi_{\rm PAH}$}
\label{appen:variation}

\begin{figure}
\plotone{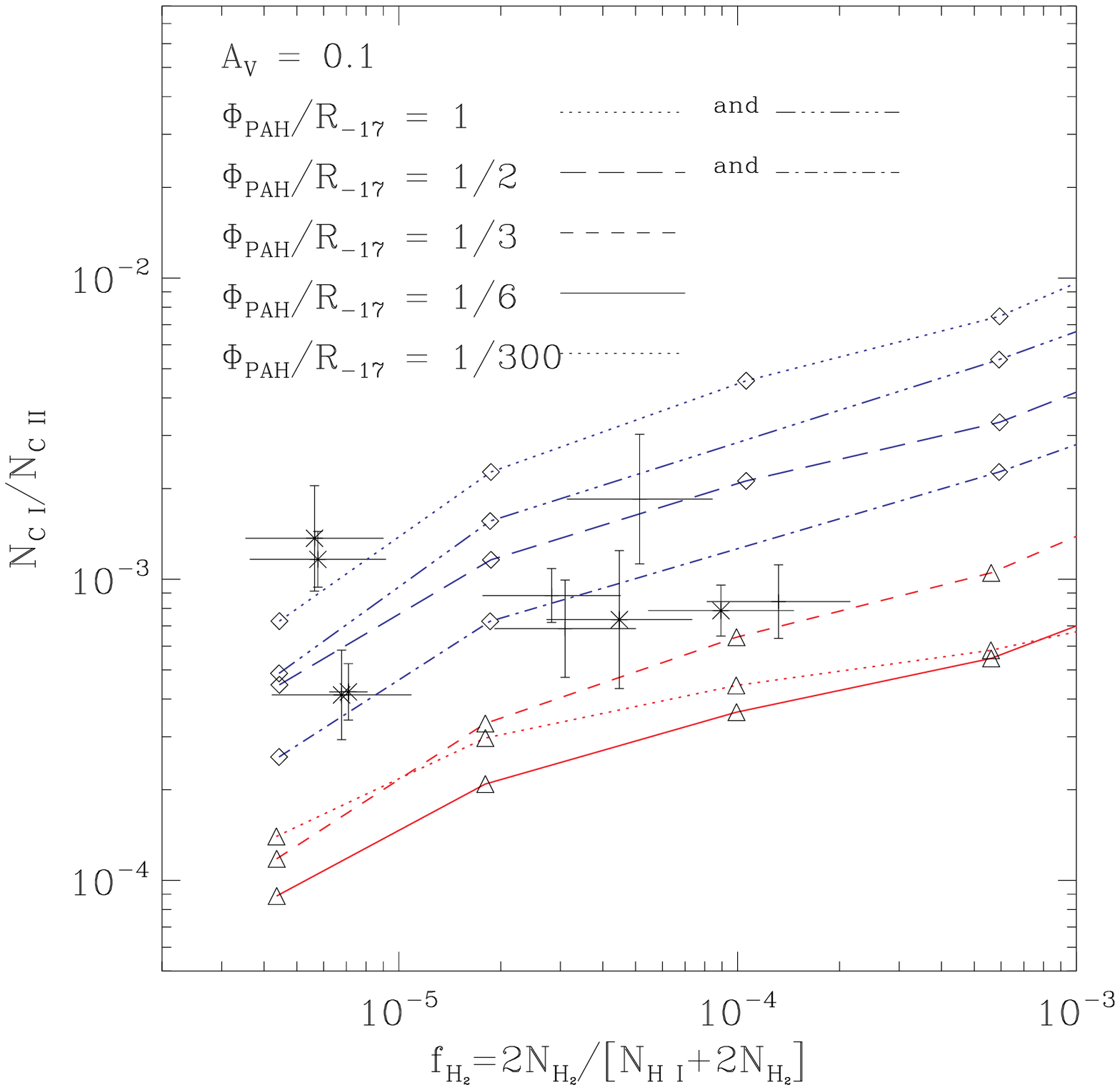}
\caption{
Column density ratio $\cicol/\ciicol$ versus molecular 
hydrogen fraction $f_{\rm H_2} = 2N_{\rm H_2}/[\hicol + 2N_{\rm
  H_2}]$. Observations are shown for cloud column densities
in the range $0.03 \la A_{\rm V} \la 0.25$, $N_{\rm H_2} < 10^{17}$
cm$^{-2}$,  and molecular fraction
$f_{\rm H_2} \la 10^{-4}$.
Clouds with $A_{\rm V} < 0.1$ are indicated with an
``$\times$''.
Curves show model results for cloud column $A_{\rm V} =
0.1$, and various values of 
$\Phi_{\rm PAH}/R_{-17}$ where $R_{-17} = R/10^{-17}$.
$\Phi_{\rm PAH} = 2.0$, $R = 2\times 10^{-17}$ ${\rm cm^{3}}$ ${\rm s^{-1}}$ 
($\Phi_{\rm PAH}/R = 1$; {\em dot}),
$\Phi_{\rm PAH} = 1$, $R = 1\times 10^{-17}$ ${\rm cm^{3}}$ ${\rm s^{-1}}$
($\Phi_{\rm PAH}/R = 1$; {\em dash-dot-dot}),
$\Phi_{\rm PAH} = 0.5$, $R = 1\times 10^{-17}$ ${\rm cm^{3}}$ ${\rm s^{-1}}$ 
($\Phi_{\rm PAH}/R = 1/2$; {\em long dash}),
$\Phi_{\rm PAH} = 1$, $R = 2\times 10^{-17}$ ${\rm cm^{3}}$ ${\rm s^{-1}}$
($\Phi_{\rm PAH}/R = 1/2$; {\em dash-dot}),
$\Phi_{\rm PAH}=1$, $R = 3\times 10^{-17}$ ${\rm cm^{3}}$ ${\rm s^{-1}}$ 
($\Phi_{\rm PAH}/R = 1/3$; 
{\em short dash})), $\Phi_{\rm PAH}=0.5$, 
$R = 3\times 10^{-17}$ ${\rm cm^{3}}$ ${\rm s^{-1}}$
($\Phi_{\rm PAH}/R = 1/6$; {\em solid}), and
$\Phi_{\rm PAH}=0.01$, $R = 3\times 10^{-17}$ ${\rm cm^{3}}$ ${\rm s^{-1}}$ 
($\Phi_{\rm PAH}/R = 1/300$; {\em dot}; lower curves).
The ratio $G_0/n$ varies along each model curve 
with higher 
$G_0/n$ yielding smaller values of $f_{\rm H_2}$.   
Individual models are shown with $n = 10$, 20, 30, and 40
${\rm cm^{-3}}$
(with the exception of the {\em dash-dot} and {\em dash-dot-dot} curves where 
$n=5$, 10, and 20 ${\rm cm^{-3}}$)  
and $G_0 = 1.7$ ($\Diamond$; {\em Blue}) and 
$G_0 = 5.1$ ($\triangle$; {\em Red}) .
}
\label{fig:ratioav025}
\end{figure}

\begin{figure}
\plotone{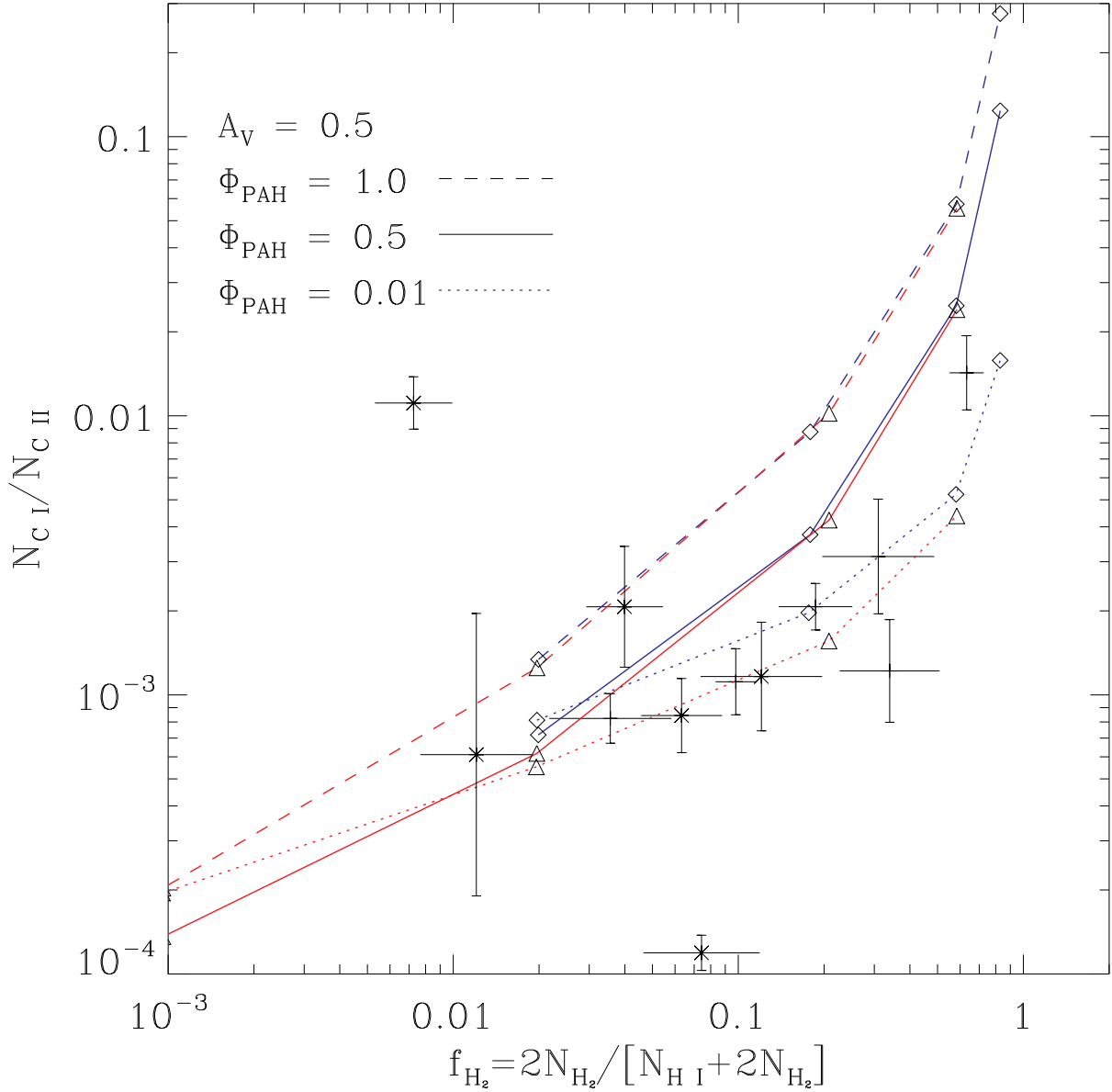}
\caption{Column density ratio $\cicol/\ciicol$ versus molecular 
hydrogen fraction $f_{\rm H_2} = 2N_{\rm H_2}/[\hicol + 2N_{\rm
  H_2}]$. Observations are shown for cloud column densities
in the range $0.25 \la A_{\rm V} \la 0.75$, and $N_{\rm H_2} > 10^{18}$ 
cm$^{-2}$.
Clouds with $A_{\rm V} < 0.5$ are indicated with an 
``$\times$''.
Curves show model results for cloud column $A_{\rm V} =
0.5$, $R=3\times 10^{-17}$ ${\rm cm^{3}}$ ${\rm s^{-1}}$,
 and $\Phi_{\rm PAH} = 1.0$ ({\em dash}), 
$\Phi_{\rm PAH} = 0.5$ ({\em solid}), and 
$\Phi_{\rm PAH} = 0.01$ ({\em dot}). The ratio
$G_0/n$ varies along each model curve 
with higher 
$G_0/n$ yielding smaller values of $f_{\rm H_2}$.   
Individual models are shown with $n = 10$, 30, 100, 
and 300 ${\rm cm^{-3}}$ and $G_0 = 1.7$ ($\Diamond$; {\em Blue}) and 
$G_0 = 5.1$ ($\triangle$; {\em Red}). For $f_{\rm H_2} = 0.02$, 
$G_0/n \sim 1.7/10$ and  for $f_{\rm H_2} = 0.2$, 
$G_0/n \sim 1.7/30$.
}
\label{fig:ratioav075}
\end{figure}

\begin{figure}
\plotone{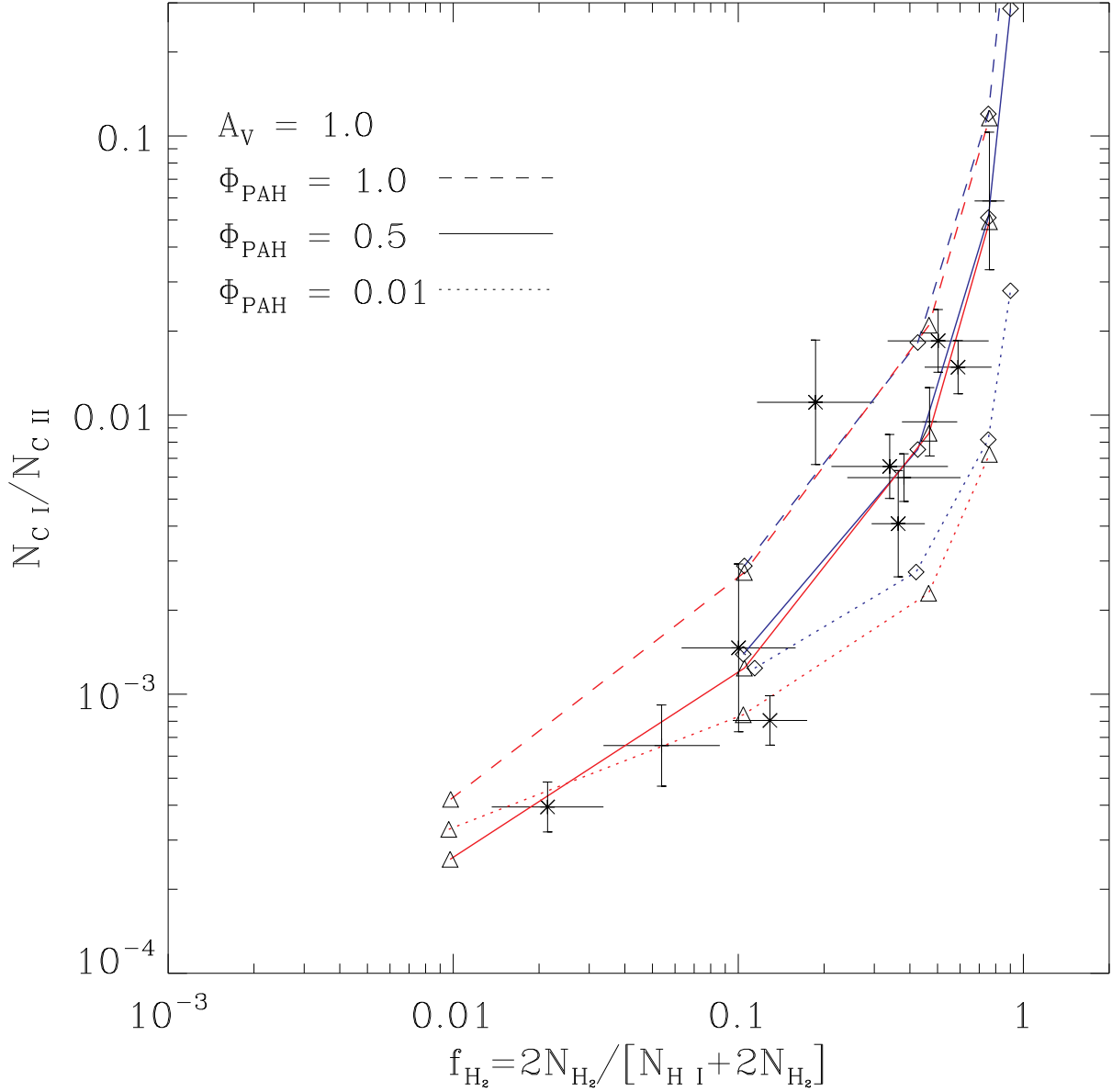}
\caption{Column density ratio $\cicol/\ciicol$ versus molecular 
hydrogen fraction $f_{\rm H_2} = 2N_{\rm H_2}/[\hicol + 2N_{\rm
  H_2}]$. Observations are shown for cloud column densities
in the range $0.75 \la A_{\rm V} \la 1.25$ and $N_{\rm H_2} > 10^{18}$
cm$^{-2}$.
Clouds with $A_{\rm V} < 1.0$ are indicated with an 
``$\times$''.
Curves show model results for cloud column $A_{\rm V} =
1.0$, $R=3\times 10^{-17}$ ${\rm cm^{3}}$ ${\rm s^{-1}}$,
 and $\Phi_{\rm PAH} = 1.0$ ({\em dash}), 
$\Phi_{\rm PAH} = 0.5$ ({\em solid}), and 
$\Phi_{\rm PAH} = 0.01$ ({\em dot}). The ratio
$G_0/n$ varies along each model curve 
with higher 
$G_0/n$ yielding smaller values of $f_{\rm H_2}$.   
Individual models are shown with $n = 10$, 30, 100, 
and 300 ${\rm cm^{-3}}$ and $G_0 = 1.7$ ($\Diamond$; {\em Blue}) and 
$G_0 = 5.1$ ($\triangle$; {\em Red}) .
}
\label{fig:ratioav125}
\end{figure}

\begin{figure}
\plotone{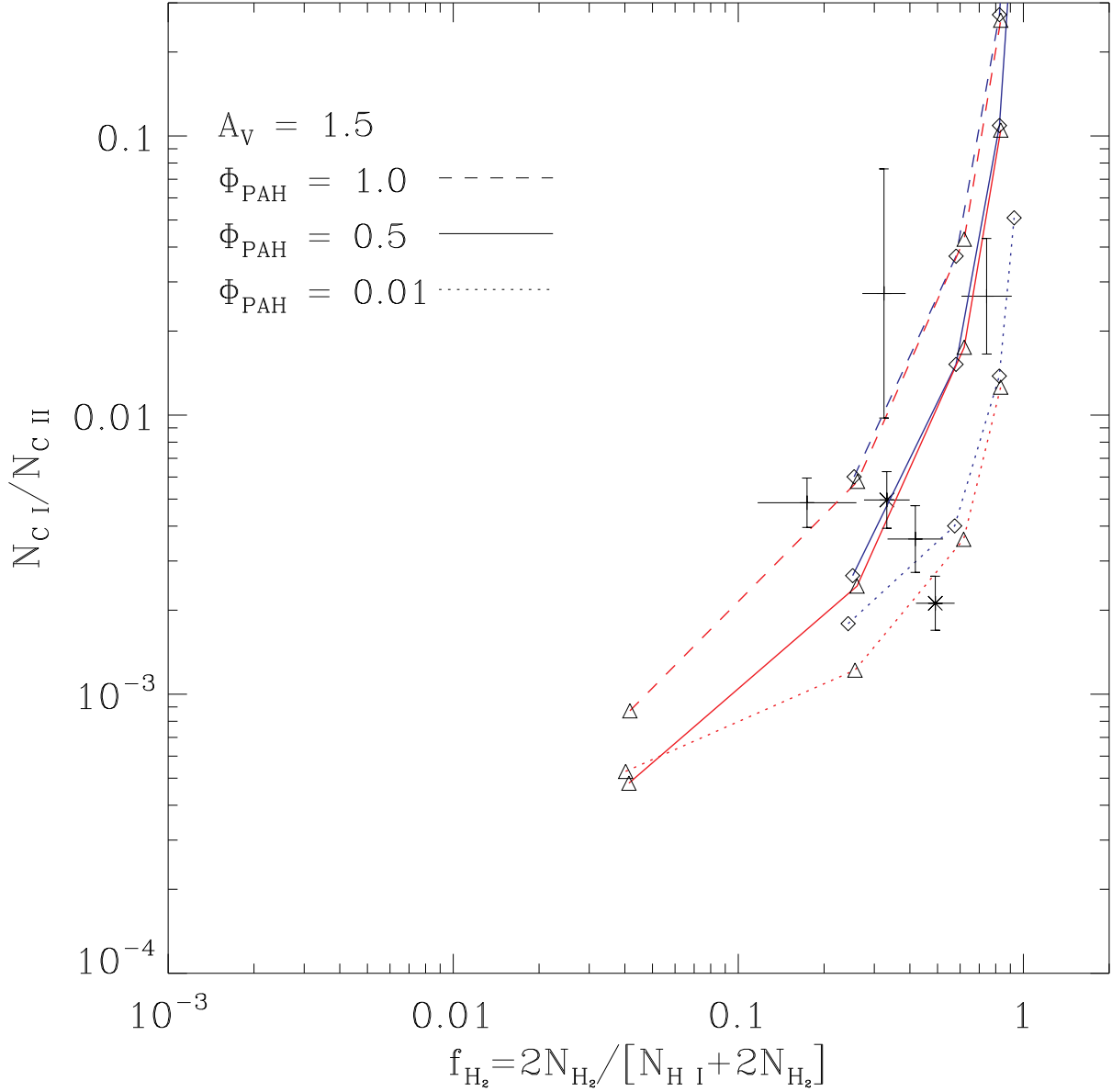}
\caption{Column density ratio $\cicol/\ciicol$ versus molecular 
hydrogen fraction 
$f_{\rm H_2} = 2N_{\rm H_2}/[\hicol + 2N_{\rm  H_2}]$. 
Observations are shown for cloud column densities
in the range $1.25 \la A_{\rm V} \la 2.13$ and $N_{\rm H_2} > 10^{18}$
cm$^{-2}$.
Clouds with $A_{\rm V} < 1.5$ are indicated with an 
``$\times$''.
Curves show model results for cloud column $A_{\rm V} =
1.5$, $R=3\times 10^{-17}$ ${\rm cm^{3}}$ ${\rm s^{-1}}$,
 and $\Phi_{\rm PAH} = 1.0$ ({\em dash}), 
$\Phi_{\rm PAH} = 0.5$ ({\em solid}), and 
$\Phi_{\rm PAH} = 0.01$ ({\em dot}). The ratio
$G_0/n$ varies along each model curve 
with higher 
$G_0/n$ yielding smaller values of $f_{\rm H_2}$.   
Individual models are shown with $n = 10$, 30, 100, 
and 300 ${\rm cm^{-3}}$ and $G_0 = 1.7$ ($\Diamond$; {\em Blue}) and 
$G_0 = 5.1$ ($\triangle$; {\em Red}) .
}
\label{fig:ratioav213}
\end{figure}

In Figures 
\ref{fig:ratioav025} through \ref{fig:ratioav213} we show the variation
in $\cicol/\ciicol$ versus $f_{\rm H_2}$ as a function of $G_0/n$ and
$\Phi_{\rm PAH}$. In the limit of low $G_0/n$ the destruction of ${\rm C^+}$
is dominated by recombination on  ${\rm PAH^-}$ and rate equation 
(\ref{eq:Arecombination}) applies. From equation (\ref{eq:Agnpahm})
we have 
\beq
       \frac{\cicol}{\ciicol} \propto \Phi_{\rm PAH}^2 
           \left [ \frac{n}{G_0} \right ]^2\,\,\, .
\label{eq:Clowgn}
\eeq
We see in Figures \ref{fig:ratioav025} through \ref{fig:ratioav213}, along
curves of constant $G_0$,  the $\cicol/\ciicol$ ratio rises steeply 
as $n$ increases.

In the limit of high $G_0/n$ the destruction of ${\rm C^+}$ is dominated
by charge exchange with ${\rm PAH^0}$ and rate equation 
(\ref{eq:Achargeexchange}) applies. From equation (\ref{eq:Agnpahn})
the $\cicol/\ciicol$ ratio is given by
\beq
       \frac{\cicol}{\ciicol} \propto \Phi_{\rm PAH}
           \left [ \frac{n}{G_0} \right ] \,\,\, ,
\label{eq:Chighgn}
\eeq
and the curves start to flatten as $n$ decreases.  

At low values of $\Phi_{\rm PAH}$ the destruction of ${\rm C^+}$ 
is dominated by gas phase recombination with electrons, $n_e$. 
In the limit of low values of $G_0/n$, the electron abundance
is determined mainly by the photoionization and recombination of 
Carbon. 
From equation   
(\ref{eq:Agnpahn}) the $\cicol/\ciicol$ ratio is given by
\beq
       \frac{\cicol}{\ciicol} \propto 
           \left [ \frac{n}{G_0} \right ] \,\,\, .
\label{eq:Clowgnlowphi}
\eeq
In the limit of high $G_0/n$, the electron abundance is mainly 
determined by the soft X-ray photoionization of H. In this case 
$n_e \propto n^{1/2}$ and
\beq
       \frac{\cicol}{\ciicol} \propto 
           \frac{n^{1/2}}{G_0} \,\,\, ,
\label{eq:Chighgnlowphi}
\eeq
and we see that the curves flatten considerably at low values of $n$.


\clearpage
\begin{deluxetable}{llcccccccl}
\rotate
\tablewidth{0pt}
\tablecaption{Observations\label{tbl:observations}}
\tablehead{
\colhead{ } &
\colhead{ } &
\colhead{$\log N_{\rm H\,I}$} &
\colhead{$\log N_{\rm H_2}$} &
\colhead{$\log N_{\rm C\,I}$} &
\colhead{$\log N_{\rm C\,II}$} &
\colhead{ } &
\colhead{ } &
\colhead{ } &
\colhead{ } \\
\multicolumn{2}{c}{Star} & 
\colhead{$({\rm cm^{-2}})$} & 
\colhead{$({\rm cm^{-2}})$} & 
\colhead{$({\rm cm^{-2}})$} & 
\colhead{$({\rm cm^{-2}})$} &
\colhead{$A_{\rm V}$}       &
\colhead{$\log f_{\rm H_2}$\tablenotemark{a}} & 
\colhead{$\log f_{\rm C\,I}$\tablenotemark{b}} &
\colhead{References} 
}
\startdata
$\kappa$ Cas & HD 2905 & $21.21\pm 0.11$ & $20.27\pm 0.18$ 
                       & $15.55\pm 0.20$ & $17.50\pm 0.22$ &
                     0.98 &   -0.73 & -1.95 &1, 2, 7\\
40 Per & HD 22951 & $21.05\pm 0.11 $ & $20.46\pm 0.18 $
                  & $15.25\pm 0.06 $ & $17.43\pm 0.11 $\tablenotemark{b} & 
                   0.85  &    -0.47 & -2.18 
                  &1, 2, 3 \\ 
$o$ Per & HD 23180 & $20.91\pm 0.11 $ & $20.61\pm 0.15 $
                   & $15.68\pm 0.06 $ & $17.41\pm 0.11 $  & 
                    0.79  &  -0.30 & -1.73 
                   &1, 2, 3 \\ 
$\zeta$ Per & HD 24398 & $20.81\pm 0.04 $ & $20.67\pm 0.10 $
                   & $15.52\pm 0.05 $ & $17.35\pm 0.08 $\tablenotemark{c}& 
                     0.79 & -0.23 & -1.83 
                   &3, 4, 5, 6 \\ 
$\chi$ Per  & HD 24534 & $20.72\pm 0.06$ & $20.92\pm 0.04$
                   &  $16.30\pm 0.20$ & $17.53\pm 0.14$\tablenotemark{c} 
                  &  1.10       & -0.12 & -1.24 & 6, 17, 22, 24\\
$\xi$ Per & HD 24912 & $21.07\pm 0.06 $ & $20.53\pm 0.08 $
                   & $15.12\pm 0.14 $ & $17.51\pm 0.13 $\tablenotemark{c}& 
                     0.98 & -0.44 & -2.39 
                   &3, 4, 6, 7 \\ 
$\epsilon$ Per & HD 24760 & $20.42\pm 0.06 $ & $19.53\pm 0.15 $
                   & $13.71\pm 0.05 $ & $16.72\pm 0.07 $  & 
                      0.16 &  -0.69 & -3.01 
                   &4, 8 \\ 
               & HD 34078 & $21.43\pm 0,10$ & $20.81\pm 0.03$ 
                   & $16.24\pm 0.44$ & $17.80\pm 0.45$ 
                   & 2.00 & -0.49 & -1.56 & 27\\
23 Ori & HD 35149 & $20.74\pm 0.08 $ & $18.30\pm 0.11 $
                   & $14.99\pm 0.05 $ & $16.94\pm 0.09 $  & 
                      0.28 & -2.14 & -1.95 
                   &4, 9, 10 \\ 
$\delta$ Ori & HD 36486 & $20.19\pm 0.03 $ & $14.74\pm 0.05 $
                   & $13.02\pm 0.09 $ & $16.39\pm 0.09 $   & 
                    0.085 & -5.15 & -3.37 
                   &7, 11, 12 \\ 
$\lambda$ Ori & HD 36861 & $20.79\pm 0.08 $ & $19.11\pm 0.11 $
                   & $14.32\pm 0.04 $ & $17.00\pm 0.21 $\tablenotemark{c} &
                   0.32 &  -1.40 & -2.68 
                   &4, 6, 8 \\ 
$\iota$ Ori & HD 37043 & $20.16\pm 0.05 $ & $14.69\pm 0.20 $\tablenotemark{d}
                   & $12.98\pm 0.14 $ & $16.36\pm 0.15 $ & 
                   0.071 & -5.17 & -3.38 
                   &2, 4, 7 \\ 
$\epsilon$ Ori & HD 37128 & $20.46\pm 0.07$ & $16.28\pm 0.20 $       
                   & $13.59\pm 0.10$ & $16.66\pm 0.12$ & 
                   0.14 & -3.88 & -3.07  & 4, 7, 12\\
$\zeta$ Ori  &  HD 37742 & $20.41\pm 0.08 $ & $15.82\pm 0.20$ 
                   & $13.88\pm 0.20$ & $16.61\pm 0.22$ &
                   0.13 & -4.29 & -2.73 & 1, 9, 10, 13\\
$\mu$ Col & HD 38666 & $19.86\pm 0.08$ & $15.51\pm 0.20$\tablenotemark{d}
                   & $12.96\pm 0.03$ & $16.06\pm 0.08$ &
                   0.036 &   -4.05   &  -3.10  & 2, 23\\
$\kappa$ Ori & HD 38771 & $20.53\pm 0.04$ & $15.68\pm 0.20$\tablenotemark{d} 
                   & $13.68\pm 0.08$ & $16.73\pm 0.09$ &
                   0.17 & -4.55 & -3.05 & 2, 4, 7\\                   
139 $\tau$   &  HD 40111& $20.90\pm 0.08$ & $19.74\pm 0.20$
                   & $14.23\pm 0.18$ & $17.16\pm 0.19$ &
                   0.46 & -0.92 & -2.93 & 1, 2, 7 \\
15 Mon      & HD 47839 & $20.36\pm 0.06$ & $15.55\pm 0.20$\tablenotemark{d} 
                   & $13.40\pm 0.15$ & $16.56\pm 0.16$ &
                   0.13 & -4.51 & -3.16 & 4, 2, 14\\
$\zeta$ Pup & HD 66811 & $19.99\pm 0.02$ & $14.45\pm 0.20$\tablenotemark{d}
                   & $13.26\pm 0.09$ & $16.19\pm 0.09$ &
                   0.049 & -5.24 & -2.93 & 1, 2, 14\\
$\gamma^2$ Vel & HD 68273 & $19.78\pm 0.04$ & $14.23\pm 0.20$\tablenotemark{d}
                   & $13.12\pm 0.17$ & $15.98\pm 0.17$ &
                   0.031 & -5.25 & -2.86 & 1, 2, 14\\
$\rho$ Leo    & HD 91316 & $20.26\pm 0.08$ & $15.61\pm 0.20$\tablenotemark{d}
                   & $13.33\pm 0.22$ & $16.46\pm 0.23$ &
                   0.091 & -4.35 & -3.13 & 1, 2, 14\\
             & HD 112244 & $21.08\pm 0.08$ & $20.14\pm 0.11$ 
                         & $14.69\pm 0.05$ & $17.37\pm 0.08$
                   & 0.74 & -0.73  & -2.68 & 1, 2, 8\\
 1 Sco       & HD 141637 & $21.19\pm 0.08$ & $19.23\pm 0.18$ 
                   & $14.00\pm 0.05$ & $17.40\pm 0.09$ &
                    0.80 & -1.67 & -3.40 & 1, 2, 8\\
$\pi$ Sco  & HD 143018 & $20.72\pm 0.04$ & $19.32\pm 0.20$ 
                   & $13.03\pm 0.05$ & $16.95\pm 0.06$ &
                      0.28 & -1.13& -3.92 & 1, 2, 8\\
$\delta$ Sco & HD 143275 & $21.14\pm 0.08$ & $19.41\pm 0.20$ 
                   & $14.25\pm 0.05$ & $17.36\pm 0.09$ &
                     0.74 &  -1.45 & -3.11 & 1, 2, 8\\
$\beta^1$ Sco & HD 144217 & $21.10\pm 0.04$ & $19.83\pm 0.06$ 
                    & $14.42\pm 0.06$ & $17.37\pm 0.10$\tablenotemark{c}&
                   0.70 & -1.01 & -2.92 & 1, 2, 6, 15\\ 
$\omega^1$ Sco & HD 144470 & $21.18\pm 0.08$ & $20.05\pm 0.11$ 
                   & $14.35\pm 0.05$ & $17.44\pm 0.09$ &
                    0.88 & -0.89 & -3.09 & 1, 2, 8 \\
$\nu$ Sco & HD 145502 & $21.14\pm 0.15$ & $19.89\pm 0.15$ 
                   & $14.56\pm 0.27$ & $17.39\pm 0.30$ &
                    0.78 & -1.00 & -2.83& 1, 2, 7\\
$\sigma$ Sco & HD 147165 & $21.34\pm 0.15$ & $19.79\pm 0.15$ 
                   & $14.38\pm 0.05$ & $17.56\pm 0.15$ &
                  1.15 & -1.27 & -3.18 & 1, 2, 8\\
$\rho$ Oph & HD 147933A & $21.55\pm 0.10$ & $20.57\pm 0.15$
                   & $15.52\pm 0.02$ & $17.83\pm 0.09$ &
                   2.13 &-0.76 & -2.31 & 2, 15, 16\\ 
$\chi$ Oph & HD 148184 & $21.14\pm 0.08$ & $20.63\pm 0.18$ 
                   & $15.33\pm 0.01$ & $17.55\pm 0.09$ &
                   1.15 & -0.42 & -2.22 & 1, 2, 15\\
22 Sco & HD 148605    & $20.95\pm 0.08$ & $18.74\pm 0.18$ 
                    & $13.95\pm 0.50$ & $17.16\pm 0.51$ &
                     0.46 & -1.92 & -3.21 & 1, 2, 7\\
$\zeta$ Oph & HD 149757 & $20.72\pm 0.02$ & $20.65\pm 0.05$
                    & $15.51\pm 0.07$ & $17.36\pm 0.11$\tablenotemark{c} &
                     0.71 & -0.20 & -1.84 & 2, 4, 6\\
          & HD 154368  & $21.00\pm 0.05$ & $21.16\pm 0.07$
                    & $16.22\pm 0.20$ & $17.79\pm 0.21$ &
                   1.96 & -0.13 & -1.57 &  17, 18\\
$\gamma$ Ara & HD 157246 & $20.71\pm 0.06$ & $19.24\pm 0.13$ 
                    & $13.87\pm 0.12$ & $16.94\pm 0.13$ &
                   0.26 &  -1.20  & -3.07 & 4, 7\\
$\kappa$ Aql & HD 184915 & $20.90\pm 0.11$ & $20.31\pm 0.15$ 
                    & $14.37\pm 0.16$ & $17.28\pm 0.18$ &
                  0.61 &    -0.47 & -2.91 & 1, 2, 7\\
             & HD 185418   & $21.11\pm 0.15$ & $20.76\pm 0.05$
                           & $15.57\pm 0.09$ & $17.59\pm 0.12$ &
                      1.24 &   -0.33 & -2.02 &17,  25, 26 \\ 
             & HD 192639 & $21.29\pm 0.09$ & $20.69\pm 0.05$
                    & $15.37\pm 0.08$ & $17.67\pm 0.10$ &
                   1.55 & -0.48 & -2.30 & 4, 17, 20\\
59 Cyg & HD 200120 & $20.25\pm 0.08$ & $19.32\pm 0.18$ 
                    & $13.94\pm 0.15$ & $16.54\pm 0.17$ &
                   0.11 &  -0.72 & -2.60 & 1, 2, 7\\
       & HD 206267 & $21.31\pm 0.15$ & $20.86\pm 0.04$ 
                 & $15.30\pm 0.08$\tablenotemark{e} & $17.74\pm 0.12$ &
                 1.74 &   -0.38   & -2.44 & 17, 21\\
       & HD 210839 & $21.16\pm 0.10$ & $20.84\pm 0.04$ 
                    & $14.98\pm 0.08$ & $17.65\pm 0.10$ &
                   1.42 &  -0.31 & -2.67 & 17, 21, 22\\ 
$\sigma$ Cas & HD 224572 & $20.88\pm 0.08$ & $20.23\pm 0.18$
                    & $14.74\pm 0.19$ & $17.24\pm 0.21$ &
                0.55 &   -0.51 & -2.50 & 1, 2, 7\\

\enddata
\tablerefs{
(1)~\cite{bohlin1978}; (2)~\cite{savage1977}; (3)~\cite{wannier1999};
(4)~\cite{cartledge2004}; 
(5)~\cite{snow1977}; (6)~\cite{sofia2004}; (7)~\cite{jenkins1983};
(8)~\cite{zsargo2003}; (9)~\cite{welty2003}; (10)~Compilation of column densities by 
D. Welty http://astro.uchicago.edu/home/web/welty/coldens.html; (11)~\cite{jenkins1999};
(12)~\cite{jenkins2000}; (13)~\cite{jenkins1997};
(14)~\cite{jenkins1979}; (15)~\cite{zsargo1997};
(16)~\cite{shull1985}; (17)~\cite{rachford2002}; (18)~\cite{snow1996};  
(19)~\cite{gry2002}; (20)~\cite{sonn2002}; (21)~\cite{jt2001};
(22)~\cite{diplas1994}; (23)~\cite{howk1999}; (24)~\cite{snow1998};
(25)~\cite{fitz1990}; (26)~\cite{sonn2003}; (27)~\cite{boisse2005}
}
\tablenotetext{a}{$f_{\rm H_2} = 2N_{\rm H_2}/[\hicol + 2N_{\rm H_2}]$.}
\tablenotetext{b}{$f_{\rm C\,I} = \cicol/\ciicol$.} 
\tablenotetext{b}{$\ciicol$ is taken as $1.6\times 10^{-4}\times [ \hicol + 2N_{\rm H_2} ]$
   unless otherwise noted.}
\tablenotetext{c}{Observed $\ciicol$.}
\tablenotetext{d}{Uncertainties for low $N_{\rm H_2}$ columns were 
not given in \cite{savage1977}.
We use the uncertainty suggested by Welty [Reference (10)].} 
\tablenotetext{f}{\ion{C}{1} column uncertainty assumed to be similar
              to \cite{sonn2002}.}


\end{deluxetable}

\begin{deluxetable}{ll}
\tablecaption{Reaction Rates\label{tbl:reactions}}
\tablehead{
\colhead{ } &
\colhead{ Rate } \\
\colhead{Reaction} &
\colhead{(${\rm cm^{3}}$ ${\rm s^{-1}}$)} 
}
\startdata
${\rm PAH^-} + {\rm H^+} \rightarrow {\rm PAH^0} + {\rm H}$
   & $8.3\times 10^{-7} \Phi_{\rm PAH} T_2^{-0.5}$ \tablenotemark{a}\\
${\rm PAH^0} + {\rm H^+} \rightarrow {\rm PAH^+} + {\rm H}$
   & $3.1\times 10^{-8} \Phi_{\rm PAH}$ \\
${\rm PAH^+} + e \rightarrow {\rm PAH^0}$
   & $3.5\times 10^{-5} \Phi_{\rm PAH} T_2^{-0.5}$\\
${\rm PAH^0} + e \rightarrow {\rm PAH^-} $
   & $1.3\times 10^{-6} \Phi_{\rm PAH}$ \\
${\rm PAH^-} +{\rm C^+} \rightarrow {\rm PAH^0} + {\rm C}$ 
   &$2.4\times 10^{-7}\Phi_{\rm PAH} T_2^{-0.5}$ \tablenotemark{b} \\
${\rm PAH^0} + {\rm C^+}\rightarrow {\rm PAH^+} + {\rm C}$ 
    & $8.8\times 10^{-9}\Phi_{\rm PAH}$  \tablenotemark{b} \\
${\rm PAH^0} + h\nu \rightarrow {\rm PAH^+} + e$ &
         $4.6\times 10^{-9}G_0\exp (-1.8 A_{\rm V})$ \tablenotemark{c}\\
${\rm PAH^-} + h\nu \rightarrow {\rm PAH^0} + e$ &
         $1.2\times 10^{-8}G_0\exp (-1.8 A_{\rm V})$ \tablenotemark{c}\\
${\rm C} + h\nu \rightarrow {\rm C^+} + e$ &
         $2.1\times 10^{-10}G_0\exp (-2.6 A_{\rm V})$ \tablenotemark{c}\\
${\rm C^+} + e\rightarrow {\rm C} + h\nu$ &
         $1.8\times 10^{-11}T_2^{-0.83}$ \\
\enddata
\tablenotetext{a}{$T_2 = T/100$ K.}
\tablenotetext{b}{Additional collisonal rates scale as $(m)^{-0.5}$
 where $m$ is the mass of the collision partner.} 
\tablenotetext{c}{Photo rates have units ${\rm s^{-1}}$.
 $G_0$ is the FUV field measured in units of the \cite{habing1968}
field ($=1.6\times 10^{-4}$ erg cm$^{-2}$ s$^{-1}$ sr$^{-1}$). The
\cite{draine1978} field is a factor of 1.7 larger. }
\end{deluxetable}

\clearpage

\end{document}